\documentclass[fleqn,usenatbib]{mnras}

\usepackage{newtxtext,newtxmath}

\usepackage[T1]{fontenc}

\DeclareRobustCommand{\VAN}[3]{#2}
\let\VANthebibliography\thebibliography
\def\thebibliography{\DeclareRobustCommand{\VAN}[3]{##3}\VANthebibliography}

\usepackage{graphicx}	
\usepackage{amsmath}	

\usepackage{pdflscape}
\usepackage{xspace}
\usepackage{xcolor}
\usepackage{float}
\usepackage[normalem]{ulem}

\def\kms{{\rm km\,s$^{-1}$}\xspace}

\def\Rsun{{\rm R$_{\odot}$}\xspace}
\def\Msun{{\rm M$_{\odot}$}\xspace}
\def\arcsec{^{\prime\prime}}
\newcommand{\io}[2]{#1\,{\textsc{#2}}}

\title[Searching for Galactic LRN]{Searching for the Next Galactic Luminous Red Nova}

\author[H. Addison et al.]{
\parbox{17.5cm}{
Harry Addison$^{1,2}$\thanks{addisonh@sky.com},
Nadejda Blagorodnova$^{2}$,
Paul J. Groot$^{2,3,6}$,
Nicolas Erasmus$^{3}$,
David Jones$^{4,5}$,
Orapeleng Mogawana$^{3,6}$
}
\\
\\
$^{1}$Department of Physics, University of Surrey, Guildford GU2 7XH, United Kingdom\\
$^{2}$Department of Astrophysics/IMAPP, Radboud University, P.O. Box 9010, 6500 GL Nijmegen The Netherlands\\
$^{3}$South African Astronomical Observatory, PO Box 9, Observatory 7935, Cape Town, South Africa\\
$^{4}$Instituto de Astrof\'isica de Canarias, E-38205 La Laguna, Tenerife, Spain\\
$^{5}$Departamento de Astrof\'isica, Universidad de La Laguna, E-38206 La Laguna, Tenerife, Spain\\
$^{6}$University of Cape Town, Private Bag X3, Rondebosch 7701, Republic of South Africa\\
}

\date{Accepted XXX. Received YYY; in original form ZZZ}

\pubyear{2022}

\begin{document}
\label{firstpage}
\pagerange{\pageref{firstpage}--\pageref{lastpage}}
\maketitle

\begin{abstract} 
Luminous red novae (LRNe) are astrophysical transients believed to be caused by the partial ejection of a binary star's common envelope (CE) and the merger of its components. The formation of the CE is likely to occur during unstable mass transfer, initiated by a primary star which is evolving off the main sequence (a Hertzsprung gap star) and a lower mass companion.
In agreement with observations, theoretical studies have shown that outflows from the pre-CE phase produce a detectable brightening of the progenitor system a few years before the ejection event.
Based on these assumptions, we present a method to identify Galactic LRNe precursors, the resulting precursor candidates, and our follow-up analysis to uncover their nature.
We begin by constructing a sample of progenitor systems, i.e. Hertzsprung gap stars, by statistically modelling the density of a colour magnitude diagram formed from ``well behaved'' Gaia DR2 sources. Their time-domain evolution from the Zwicky Transient Facility (ZTF) survey is used to search for slowly brightening events, as pre-CE precursor candidates. The nature of the resulting candidates is further investigated using archival data and our own spectroscopic follow-up.
Overall, we constructed a sample of $\sim5.4\times{10^4}$ progenitor sources, from which 21 were identified as candidate LRNe precursors. Further analysis revealed 16 of our candidates to be H$\alpha$ emitters, with their spectra often suggesting hotter (albeit moderately extincted) A-type or B-type stars. Because of their long-term variability in optical and mid-infrared wavelengths, we propose that many of our candidates are mass-transferring binaries with compact companions surrounded by dusty circumstellar disks or alternatively magnetically active stellar merger remnants.

\end{abstract}



\section{Introduction}

Luminous red novae (LRNe) are astrophysical transients with a peak brightness located between that of novae and supernovae. They are thought to be caused by the partial ejection of a binary star's common envelope (CE), followed by the merger of its stellar components \citep{SokerTylenda2003ApJ,Tylenda2011,Pejcha2014, Pejcha2016,Blagorodnova_2017}. The formation of the CE is caused by dynamically unstable mass transfer from the binary's primary star to its secondary companion \citep{Paczynski1976}. LRNe  progenitor studies suggest that this usually occurs when the primary star starts to evolve off the main sequence (MS) towards the red giant branch (RGB) \citep{MacLeod_2017,Blagorodnova_2017,Blagorodnova_2021}. During this short-lived stage, the star becomes a yellow giant (YG) or a yellow super giant (YSG), and quickly crosses the so-called Hertzsprung gap (HG), when the radius of the star grows by nearly an order of magnitude on timescales of only a few thousand years.

In binary systems with a small enough separations, the fast expansion of the primary component is likely to lead to a Roche lobe overflow (RLOF) through the inner Lagrange point, L1, initiated by the thermal-timescale expansion of the radiative envelope. As mass is lost through the overflow, the primary cannot maintain hydrostatic equilibrium in the envelope, which accelerates the mass-transfer rate, leading to a runaway and causing the formation of the CE \citep{MacLeod2020ApJ_runaway}. This initial stage of mass transfer is predicted to play an important role in the removal of mass and angular momentum from the system, as part of the gas escapes via the outer Lagrange potential points, L2 and L3  \citep{Pejcha2014,Pejcha2016,Pejcha2017,MacLeod2020ApJ_premasslossM}. Another possible explanation would be the existence of jets powered by enhanced accretion onto the companion \citep{Soker2015ApJ, soker_2020, soker_2021}.

Outflows during the pre-CE phase have also shown to cause a brightening of the progenitor system just a few years before the CE ejection. This precursor emission has been consistently observed in archival observations of LRNe shortly before their main outbursts. Examples of this include V1309 Scorpii \citep{Tylenda2011}, SNHunt248 \citep{Kankare2015AA}, M101 OT2015-1 \citep{Blagorodnova_2017}, and M31-LRN-2015 \citep{blagorodnova2020}, where these systems brightened by 1$-$3 magnitudes within the last $\sim$5\,years before the LRN event. The interaction of the dynamical ejecta with the mass lost during the precursor phase can possibly explain the observed diversity of LRN light curves \citep{metzger2017,MatsumotoMetzger2022arXiv}.

Currently, the study of LRNe precursors has only relied on observations retrieved from archival data, which are usually scattered and rarely include multi-wavelength information. Having the ability to identify LRNe ahead of time would enable a much more detailed analysis of the progenitor system using high cadence multi-wavelength observations and spectroscopic studies. These observations would allow us to better understand the characteristics of the progenitor system, and shed light on the mass transfer mechanisms leading to the final CE ejection. As an example, the detailed $\sim$8\,year long pre-outburst archival observations of the Galactic LRN V1309\,Sco \citep{Tylenda2011} have already revolutionized CE research \citep{Pejcha2014,Nandez2014ApJ,MacLeod2020ApJ_premasslossM,Hatfull2021MNRAS}. However, if identified a few years before its outburst phase, additional observations of the system could have provided crucial data on the mass loss and outflows from the system shortly before its final merger. With an expected Galactic rate of LRNe between 0.5 and 0.03 yr$^{-1}$ for low-luminosity V4332\,Sgr- and brighter V838\,Mon-like objects \citep{Kochanek2014}, the odds of finding the next LRN precursor are favourable.

This paper presents a novel method for identifying Galactic LRNe candidates, which are likely to outburst within the next 1$-$10\,years. Previous works aiming to identify binaries on their path towards an imminent merger, relied on the identification of eclipsing binaries with a noticeable period decay, similar to the one experienced by V1309\,Sco \citep{Pietrukowicz2017AcA,Gazeas2021MNRAS,Hong2022AJ}. However, these searches are usually biased towards low-mass main-sequence companions, where the radius difference between primary and secondary star is relatively small. As a complement to these works, our method is focused on the detection of precursor emission from more evolved stars located in the HG, with masses $M>$2\,\Msun. This parameter space is more representative of the LRNe progenitor systems that have been observed at extra-galactic distances, and therefore serves as a pioneering proof of concept for searches outside of the Milky Way with ongoing and future time-domain facilities, such as ZTF \citep{Bellm_2018, Masci_2018}, MeerLICHT/BlackGEM \citep{Bloemen_2016, Groot_2019}, and LSST \citep{Ivezic_2019}.

This paper is organized as follows: in Section \ref{sec:Method} we describe our method to identify LRNe progenitors, and search for the precursors. The results of our search are presented in Section \ref{sec:Results}. In Section \ref{sec:Discussion} we provide an assessment of the limitations and future improvements for our methodology, and discuss the nature of the selected candidate systems. Finally, in Section \ref{sec:Conclusion} we summarise the conclusions of the study.

\section{Selection strategy}\label{sec:Method}

Our method to first identify likely LRNe progenitor systems uses the {\sl Gaia} DR2 and EDR3 catalogues \citep{Gaia_DR2_2018,Gaia_EDR3_2021}, and then searches for sustained brightening activity using time-domain data from the {\sl Zwicky Transient Facility} (ZTF) time-domain survey \citep{ZTF_DR7}. Our method is complemented with archival analysis (see Sect. \ref{sec:archival}) and follow-up observations for 21 candidate systems (Sect. \ref{sec:followup}).

\subsection{Progenitor Selection}\label{sec:Progenitor_Selection_Method}

The initial stage of our method focuses on the identification of binaries that contain a YG or YSG primary component that has finished hydrogen core burning. The initiation of hydrogen shell-burning at the base of the envelope quickly expands the stellar radius towards the red giant branch (RGB). If a close companion is present with orbital separations, $a$ smaller than $\sim$2 R$_{RGB}$ (neglecting mass losses from the system), the growth of the stellar radius on thermal timescales initiates an unstable (case B) mass transfer on to the companion, which can result in the formation of a CE. The YSG and YG phases have time spans of around 1 per cent of the main sequence phase \citep{Drout_2009}, which implies that the population of such sources is small. Therefore, they reside in a thinly populated region of the Hertzsprung Russell diagram (HRD), known as the HG. Our progenitor identification method exploits this knowledge to select an initial set of likely progenitors.

Previous studies by \citet{Drout_2009, Neugent_2010, Drout_2012, Neugent_2012} have identified YSG populations in fields containing the nearby galaxies M31, the Small Magellanic Cloud, M33, and the Large Magellanic Cloud. Their approach was to select F- and G-type stars (YSGs and foreground dwarf star contaminants) from the CMDs of the fields by applying suitable cuts. Nevertheless, such samples had an important source of contamination related from low-mass foreground stars, which was mitigated with additional spectroscopic observations of the candidates, as the large systemic velocities of M31 and the SMC allowed them to distinguish extragalactic supergiants based on their radial velocities. In our work, the Gaia EDR3 already contains the parallaxes of the sources, which allows us to remove contamination from foreground stars efficiently. 

To define the location of the HG, we choose to initially model the observed density of stars in the Gaia CMD. The HG in this model is a low density region located between the denser MS and RGB regions. Among the possible density estimation techniques, we chose the Gaussian mixture model \citep[GMM;][]{Pearson_1984, McLachlan_2019} because of its straightforward interpretation of probabilities. This modelling method uses a set of 2D Gaussian distributions which are fit to a training set in order to represent the density of the data. With the model formed, another data set can then be fitted with this model to make inferences about individual points. Thus, it can be inferred if a given data point belongs to the model (MS or RGB) or not. The following sections present the details of the model and the data sets used.

\begin{table*}
    \caption{Constraints and ADQL queries used to obtain the training data set from Gaia DR2 \citep{Gaia_DR2_2018}. The constraints placed on the parallax over error, flux over error, RUWE, and extinction are used to provide a ``clean" sample of Galactic sources. The final constraint on the absolute $G$ magnitude is to limit the data to the region of the Hertzsprung Russell diagram that contains the Hertzsprung gap and hence likely progenitors.}
    \label{tab:Training_Data_Constraints}
    
    \centering
    \begin{tabular}{ccc}
         \hline
         \hline
         Quantity & Constraint & ADQL Query \\
         \hline
         \hline
         Parallax / Error & $\geq 20$ & gaiadr2.gaia\_source.parallax\_over\_error $\geq 20$ \\
         
         Flux / Error & $\geq 70$ & gaiadr2.gaia\_source.phot\_g\_mean\_flux\_over\_error $\geq 70$ \\
         & & gaiadr2.gaia\_source.phot\_rp\_mean\_flux\_over\_error $\geq 70$ \\
         
         RUWE & $\geq 0.95$ & gaiadr2.ruwe.ruwe  $\geq 0.95$ \\
         & $ \leq 1.05$ & gaiadr2.ruwe.ruwe $\leq 1.05$ \\
         
         Extinction & $\leq 0.75$ & gaiadr2.gaia\_source.a\_g\_val $\leq 0.75$ \\
         
         Absolute $G$ Magnitude & $\leq 1.5$ & (gaiadr2.gaia\_source.phot\_g\_mean\_mag - \\
         & & gaiadr2.gaia\_source.a\_g\_val - 10 + \\
         & & 5 * log10(gaiadr2.gaia\_source.parallax)) $\leq 1.5$ \\
         
         \hline
         \hline
         
    \end{tabular}
\end{table*}

\subsubsection{Training data set}\label{sec:Training_Data}

\begin{figure}
    \centering
    \includegraphics[width=\linewidth]{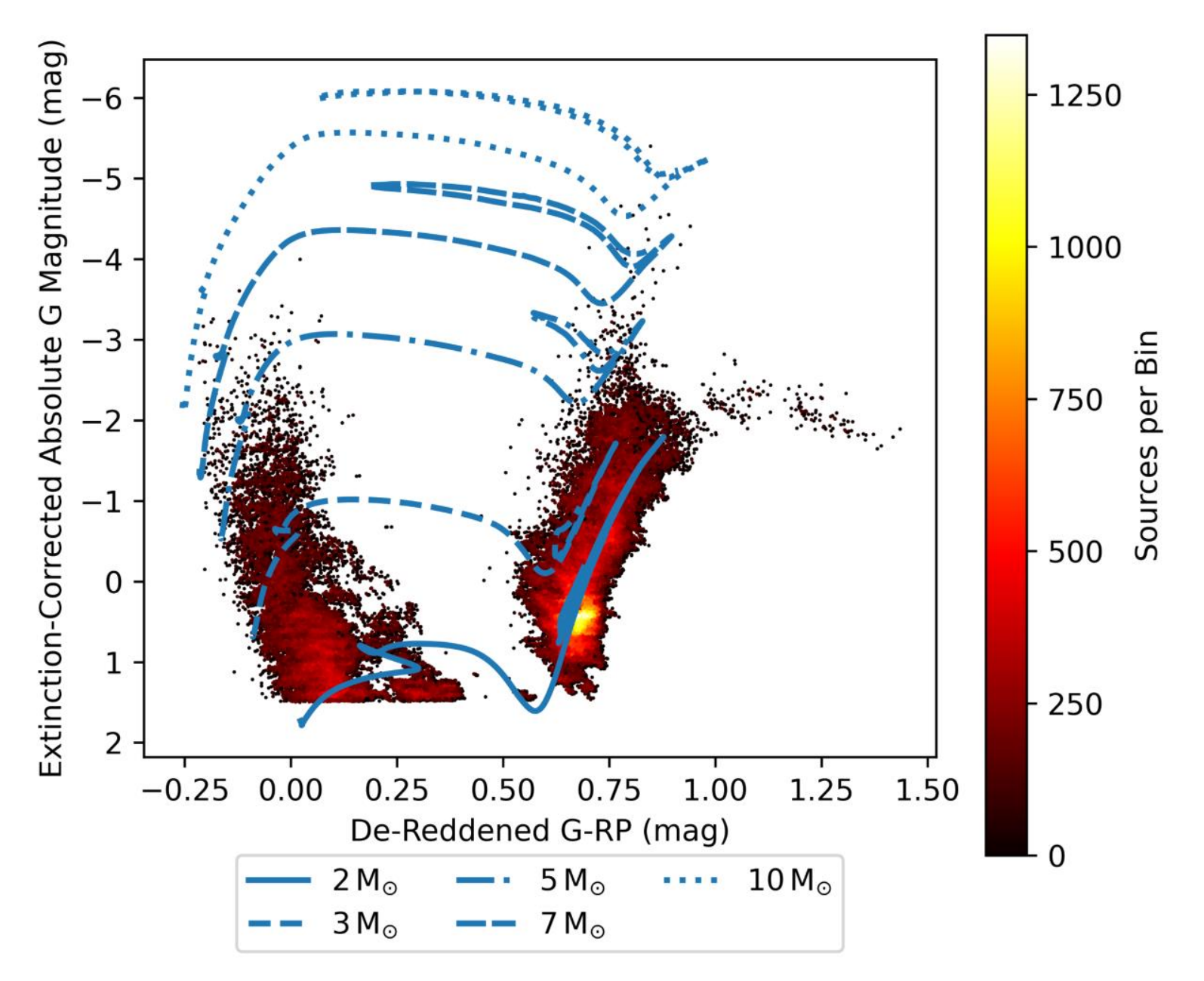}
    \caption{Training data collected from Gaia DR2 \citep{Gaia2016, Gaia_DR2_2018} presented as an extinction-corrected colour magnitude diagram with the MIST stellar evolution tracks \citep{MIST_0, MIST_1,paxton_2011, paxton_2013, paxton_2015, paxton_2018} from the MS to core helium burning phase for solar metallicity stars of masses $2$, $3$, $5$, $7$ and $10\,M_\odot$ with an initial/critical rotational velocity of 0.4.}
    \label{fig:Training_Data_CMD}
\end{figure}

The data set used to train the GMM was obtained from the Gaia data release 2 (DR2) catalogue \citep{Gaia2016, Gaia_DR2_2018}. We chose to use Gaia DR2 over Gaia EDR3 \citep{Gaia_EDR3_2021}, as it contains extinction values for individual sources, which are required to obtain an extinction-free CMD. The data set was acquired from the the online Gaia Archive \citep{Gaia_Archive}, and consisted of the following columns: apparent $G$ magnitude, $G-RP$, parallax, and $A(G)$ extinction.
Furthermore, the constraints given in Table \ref{tab:Training_Data_Constraints} were applied to the data to produce a ``clean sample'' of Galactic sources in the parameter space containing the HG. A clean sample in this case meaning that the sources have relatively low parallax errors ($\leq5$ per cent), photometric errors ($\leq1.43$ per cent), extinction values ($\leq0.75\,\rm{mag}$), and reliable astrometry, which is measured using the renormalised unit weight error (RUWE). A lower absolute $G$ magnitude limit of $1.5\,\rm{mag}$ was used to avoid contamination from lower MS stars with masses $<\sim2\,M_\odot$.

The extinction-corrected CMD containing the training data set is shown in Fig. \ref{fig:Training_Data_CMD} along with stellar evolution tracks from Modules for Experiments in Stellar Astrophysics' \citep[MESA;][]{paxton_2011, paxton_2013, paxton_2015, paxton_2018} Isochrones and Stellar Tracks \citep[MIST;][]{MIST_0,MIST_1}.  These tracks contain the evolution for single stars from MS to core helium burning phase for solar metallicity stars from $2$ to $10\,\rm{M_{\odot}}$ with an initial/critical rotational velocity of 0.4. The caveat is that the LRN progenitor systems are not single stars, but mass-transferring binaries. Because the luminosity of a mass transferring primary can drop up to two factors of magnitude before the onset of the CE \citep{Klencki2021AA}, we can not guarantee that the parameter space selected for our training and fitting data sets will be complete for primaries with masses $M_1$$\geq$2\,\Msun.

\subsubsection{Fitting data set}\label{sec:Fitting_Data}

The second data set used in our work is the fitting data set. It contains sources from Gaia EDR3 within the same parameter space as the training dataset, but this time the constraints on the data were relaxed as to include a larger number of sources. The specific ADQL queries used to select the data set are shown in Table \ref{tab:Fitting_Data_Constraints}. As before, the data set was acquired from the online Gaia Archive \citep{Gaia_Archive}, and consisted of the following columns: right ascension (RA), declination (Dec), apparent $G$ magnitude, $G-RP$ colour, and geometric distances from \citet{Bailer_Jones_2021}.
These distances were used to convert the apparent $G$ magnitudes to absolute magnitudes rather than calculating distances using the reciprocal of the parallaxes directly from Gaia EDR3. This is due to the non-trivial calculation of distances from negative and noisy parallax measurements, which \citet{Bailer_Jones_2021} overcome by applying priors. Only sources with declination higher than $-30$\,$^{\circ}$ were selected to match to the northern sky portion observed by the Zwicky Transient Facility (ZTF) time domain survey \citep{Bellm_2018, Masci_2018}, which is used at a later point in the method.

\begin{table*}
    \caption{Constraints and ADQL query conditions used to obtain the fitting data set from Gaia EDR3 \citep{Gaia_EDR3_2021}. The final two constraints on the absolute $G$ magnitude and $G-RP$ is to limit the data to the region of the Hertzsprung Russell diagram that contains the Hertzsprung gap and hence likely progenitors.}
    \label{tab:Fitting_Data_Constraints}
    
    \centering
    \small
    \begin{tabular}{ccc}
         \hline
         \hline
         Quantity & Constraint & ADQL Query \\
         \hline
         \hline
         Parallax / Error & $\geq 5$ & gaiaedr3.gaia\_source.parallax\_over\_error $\geq 5$ \\
         
         Flux / Error & $\geq 10$ & gaiaedr3.gaia\_source.phot\_g\_mean\_flux\_over\_error $\geq 10$ \\
         & & gaiaedr3.gaia\_source.phot\_rp\_mean\_flux\_over\_error $\geq 10$ \\
         
         Absolute $G$ Magnitude & $\leq 1.5$ & (gaiaedr3.gaia\_source.phot\_g\_mean\_mag - 5 * \\
         & & log10(external.gaiaedr3\_distance.r\_med\_geo / 10)) $\leq 1.5$ \\
         
         $G-RP$ & $\leq 1.22$ & gaiaedr3.gaia\_source.g\_rp $\leq 1.22$ \\
         
         Declination & $\geq -30$ & gaiaedr3.gaia\_source.dec $\geq -30$ \\
         
         \hline
         \hline
    \end{tabular}  
\end{table*}

\begin{table*}
    \caption{Parameter space cuts made to the sources in the fitting data set with log likelihoods less than the threshold value ($\rm{log}\,\mathcal{L}$ $=-6.06$). Also included are descriptions of the parameter space removed by the cuts in terms of the MS and RGB. \\
    (1) \citet{MIST_0, MIST_1,paxton_2011, paxton_2013, paxton_2015, paxton_2018}}
    \label{tab:Parameter_Space_Cuts}

    \centering
    \begin{tabular}{cc}
    
        \hline 
        \hline
        
         Cut & Parameter Space Removed \\
         
         \hline
         \hline
       
         $G \geq 1.25\rm{\,mag}$ & Data below main sequence and red giant branch. \\

         $G-RP \geq 1\rm{\,mag}$  & Data to the right of the right most \\
         & point of the red giant branch. \\
        
         $G \leq -2\rm{\,mag}$ and $G-RP \geq 0.6\rm{\,mag}$ & Data below the red giant branch. \\
         
         $G-RP \leq G-RP$ of MIST & Data to the left of the end of the main sequence \\
         main sequence ends & from the MIST  stellar evolution tracks (1). \\
         
        \hline
        \hline
    
    \end{tabular}
\end{table*}

\begin{figure}
    \centering
    \includegraphics[width=\linewidth]{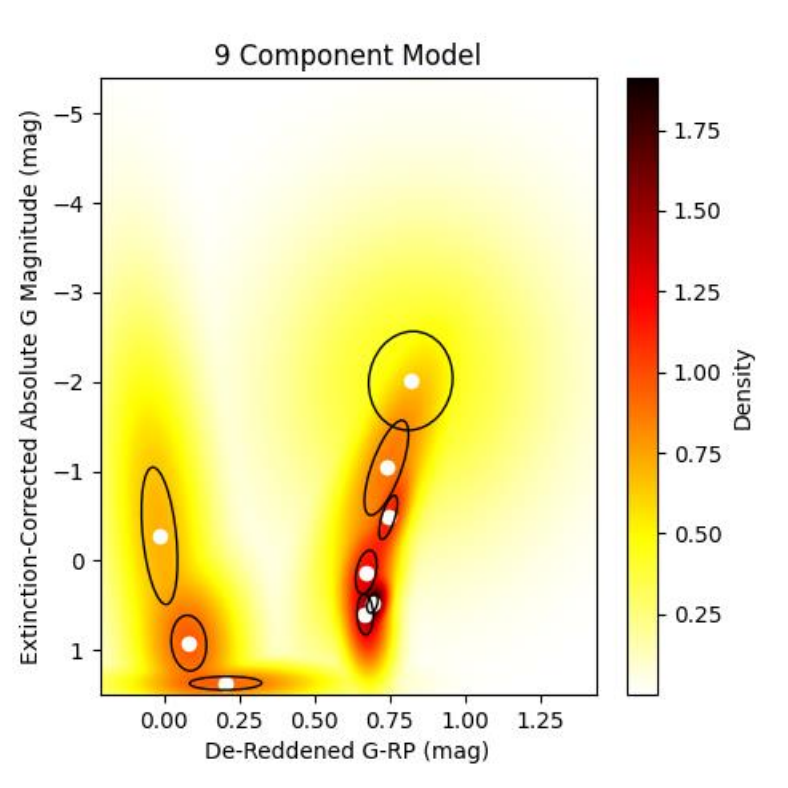}
    \caption{Nine component Gaussian mixture model of the extinction-corrected CMD constructed from the training data set. The ellipses correspond to one standard deviation of each of the two-dimensional Gaussian distributions. The collective distribution of all nine Gaussian distributions corresponds to the density distribution of the data and is shown with colour, with darker areas representing more dense regions.}
    \label{fig:9_Component_GMM}
\end{figure}

\begin{figure}
    \centering
    \includegraphics[width=\linewidth]{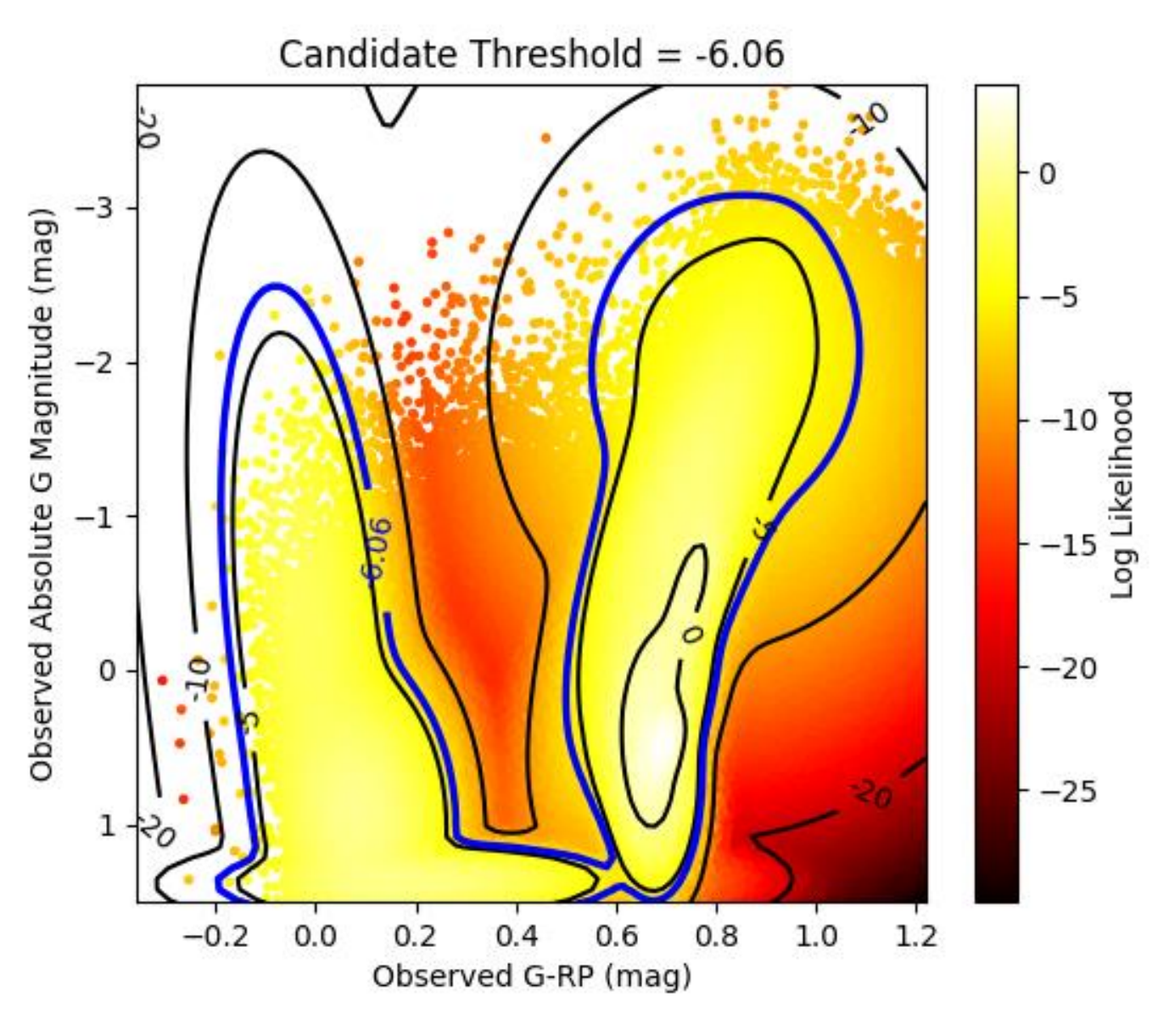}
    \caption{Observed CMD of the fitting data set with the log likelihood values of the data belonging to the GMM. The contour lines represent lines of constant log likelihood values, and the blue contour line with a value of $\rm{log}\,\mathcal{L}$ $=-6.06$ represents the log likelihood threshold used in the selection of progenitor candidates.}
    \label{fig:Fitting_Data_Probabilities}
\end{figure}

\begin{figure*}
    \centering
    \includegraphics[width=\linewidth]{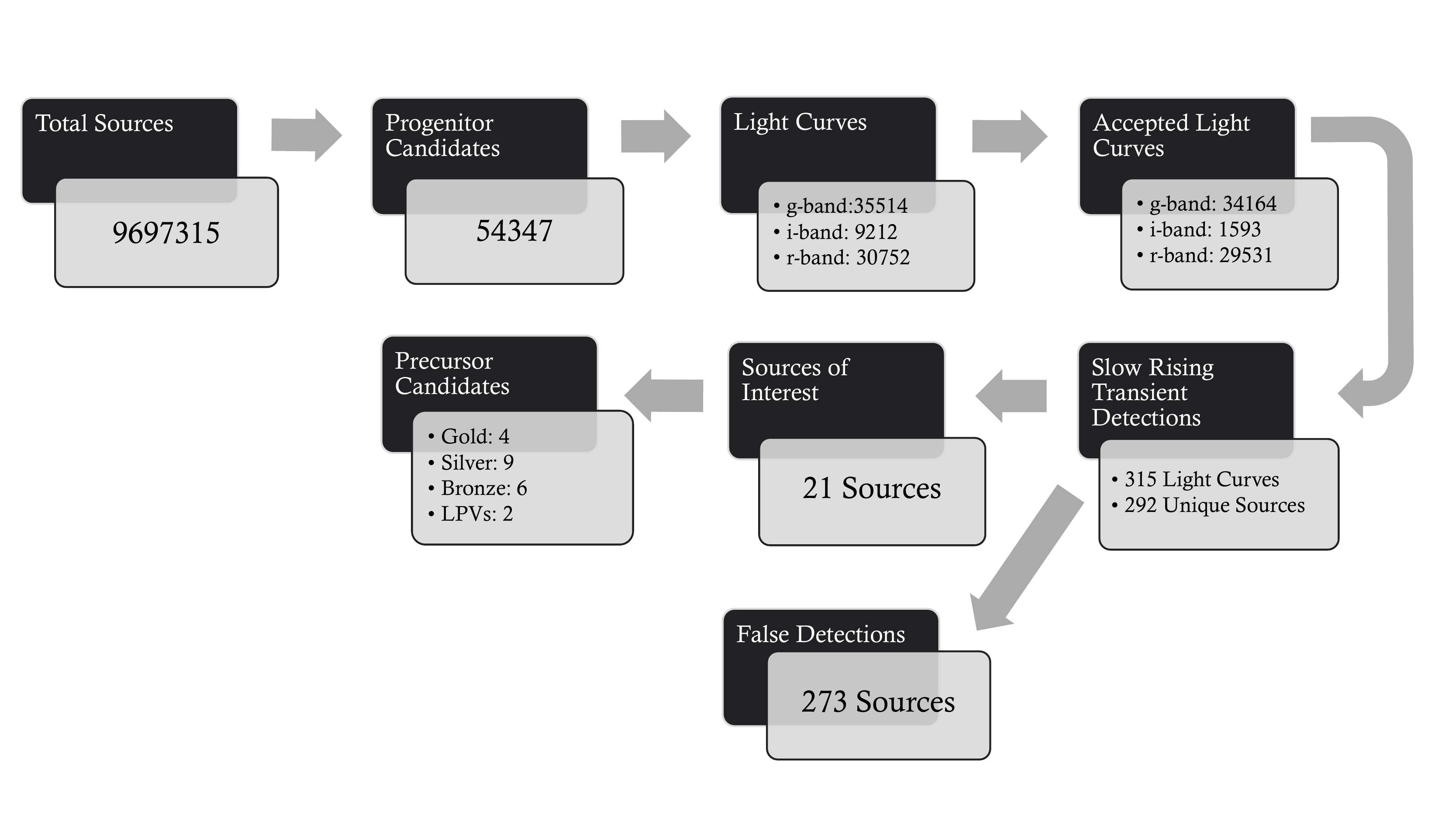}
    \caption{Overview of the number of sources at each stage of the LRN precursor selection method.}
    \label{fig:Num_Sources_Flow_Chart}
\end{figure*}

\subsubsection{Modelling the CMD density}\label{sec:CMD_Model}

A GMM was used to model the density of the extinction-corrected Galactic CMD. This modelling method was selected over other density estimation methods due to its straightforward use in assigning probabilities to belong to the model for a set of candidate sources. In our work, we used the \texttt{Python} based implementation from the \texttt{Scikit-Learn} package \citep{scikit-learn}.

To form the model of the extinction-corrected Galactic CMD, the GMM was trained using the data set described in section \ref{sec:Training_Data}. When selecting the complexity of the model, we chose to use a model that consisted of nine two-dimensional Gaussian distributions (components). This selection was based on the results from three different tests: the Bayesian information criterion \citep{Schwarz_1978}, Jenson-Shannon distance \citep{Lin_1991}, and the Silhouette \citep{Silhouette} test. These were used to compare the clustering quality and reproducibility of models with a varying number of components. The results suggested that a four-component model was best suited for the training data, however, upon visual inspection it was seen that the model density was near constant throughout the parameter space, which was not in agreement with the input training data (see Fig.\ref{fig:Training_Data_CMD}). Hence, it was apparent that a more complex model was needed. As such, the next best suitable model with nine components was selected and upon visual inspection was deemed suitable for use. The resulting GMM is presented in Fig. \ref{fig:9_Component_GMM}, where most of the density is concentrated in the MS and the RGB.

\subsubsection{Progenitor candidate selection}\label{sec:Progenitor_Candidate_Selection_Method}

To select progenitor candidates, we use our trained model to estimate the probability of containing each individual data point from our fitting data set. This probability is shown colour-coded in Fig. \ref{fig:Fitting_Data_Probabilities}, which displays the observed CMD of the fitting data set. The threshold for selecting our YG and YSG candidates was found using the cross-validation of the training data set. For each testing subset, we included as candidates all data points with a probability smaller than the model's three standard deviations. This process was repeated 100 times to mitigate possible variations due to the statistical nature of the GMM, and the resulting probabilities were averaged to give a threshold log-likelihood value of $\rm{log}\,\mathcal{L}$ $=-6.06$. This threshold log-likelihood value was then applied to the probabilities of the fitting data set. Sources with a lower probability than the threshold were selected as not belonging to the model and hence not a MS or RGB star. The next step in our method consisted in applying cuts to the parameter space to only include the region containing the HG between the MS and the RGB. The cuts made to the parameter space are provided in Table \ref{tab:Parameter_Space_Cuts}. 

The final quality cut applied to our progenitor candidates was designed to remove likely unresolved visual binaries, hence avoiding artificial light curve variations introduced by seeing. In our attempt to pair our resolution to the image quality of the ZTF survey, which has a FWHM of $2.0\arcsec$ FWHM in the $r$-band \citep{Bellm_2018, Masci_2018}, we discarded those stars from the Gaia catalogue that had a neighbouring visual companion within $2\arcsec$. To apply this condition, a $2\arcsec$ cone search centred on each of the candidate progenitors was performed in Gaia EDR3 \citep{Gaia_EDR3_2021}. Sources without neighbours were selected for our sample of YG and YSG candidates from our posterior light curve analysis to find precursors emission using time-domain data.

\subsection{Precursor Selection}\label{sec:Precursor_Selection}

Before going into outburst, the progenitor systems of LRNe have all shown a precursor emission, when the system brightened by $1-3$\,magnitudes within the prior 5 years to the optical transient. In our LRN selection process, we utilize the observational characteristics of known precursors to select new candidates among our sample of YG and YSG stars.

\subsubsection{Light curve Data}\label{sec:Light_Curve_Data}

To detect possible brightenings, we analyzed the time evolution for each source in our candidate progenitor sample. The time-domain information was obtained from the ZTF survey \citep{Bellm_2018, Masci_2018} by using the \texttt{Python} package \texttt{ztfquery} \citep{mickael_rigault_2018_1345222}. We performed cone searches with a radius of 2$\arcsec$ on each of the candidate's coordinates, collecting the available $g$, $i$, and $r$-band light curve data. The light curves were cleaned by removing possibly problematic observations having a non-zero quality \textsc{catflag} value. 

Because the observed precursor emission usually develops over longer timescales, we limited our light curve sample to sources with time coverage longer than one year. This condition also limits the number of false detections, as the expected brightening of a precursor over a time period of less than one year is small. Based on the average brightening of V1309\,Sco ($\sim0.8\,\rm{mag over} \sim1600\,\rm{days}$), the magnitude increase over less than a one year period is $\lesssim0.18\,\rm{mag}$, and so it would be difficult to distinguish from other variable sources.

One additional constraint applied to the light curves was that they had to contain more than 25 data points. This was used to also reduce the number of false detections as light curves with too few observations would not accurately sample the underlying variability. The value of 25 was chosen as about 51 per cent of light curves in ZTF DR7 have more than 20 observations \citep{ZTF_DR7}. Therefore, increasing the constraint further would reduce the number of available light curves below 51 per cent of the total number in ZTF DR7. Therefore, 25 was chosen as a balance between reducing the number of false positives and the number of available light curves.

\subsubsection{Precursor candidate selection}\label{sec:Precursor_Selection_Method}

Among all the ZTF $g$, $i$, and $r$-band light curves, we searched for precursor emission applying a slow rising transient detection method. The method initially finds the general magnitude trend of the light curve and removes any small period variations by calculating the rolling average of the magnitudes. We used a data point based bin size ($n_{bin}$) equal to a third of the total number of data points in the light curve. The rolling average was calculated using equation \ref{formula:Rolling_Average}. Next, the net change of the general trend was computed by summing the differences between consecutive rolling average values, as shown by equation \ref{formula:Sum_Differences}. This net change was then averaged over the total time span of the light curve to give an average rate of magnitude change of the source.

\begin{equation}
    n_{bin} = \frac{1}{3} \times n_{total}
\end{equation}

\begin{equation}\label{formula:Rolling_Average}
    \xi_{i} = \frac{\sum_{j=0}^{n_{bin}-1}  \left(  x_{i+j}  \right)}{n_{bin}}
\end{equation}

\begin{equation}\label{formula:Sum_Differences}
    \Theta = \sum_{i=1}^{N-1}  \left(  \xi_{i+1} - \xi_{i}   \right)
\end{equation}

Where $n_{bin}$ is the number of data points in each rolling average bin, $n_{total}$ is the total number of data points in the light curve, $\xi_{i}$ is the value of the $i^{th}$ rolling average bin, $x$ are the magnitude measurements, $\Theta$ is the sum of differences, $N$ is the total number of rolling average bins. The number of data points in each bin, $n_{bin}$, was selected to be $1/3 ~n_{total}$ as testing showed it to be one of the more suitable values in finding the general trend of the light curves.

A positive average rate indicates that the light curve's general magnitude trend is overall decreasing in brightness, whereas a negative average rate indicates that the light curve is brightening. Therefore, only light curves with negative rates are further considered. To minimize false positives from artificial brightening caused by noisy data, we apply a threshold of $-2\times{10^{-4}}\rm{\,mag/day}$ for the selection of slow rising transients. This threshold value was selected as it reduces the number of false positive detections to $\lesssim1$ per cent. Furthermore, this threshold value includes the average brightening rate of V1309\,Sco, whilst also allowing for some uncertainty in the brightening rates. 
Light curves with an average rate less than the threshold are considered to be slow rising transients.

\renewcommand{\tabcolsep}{0.15cm}
\begin{table}
    \caption{Break down of the number of light curves of the 54347 progenitor candidates at the different stages of analysis: progenitor sources with ZTF light curves \citep{Masci_2018,Bellm_2018}, accepted light curves that satisfy the quality constraints, light curves that satisfy the brightening test, and the number of sources in the ATLAS variable star catalogue \citep{Heinze_2018}. Also noted are the percentages of sources and light curves that are present from the previous step in the analysis.}
    \label{tab:LC numbers}
    \centering
    \begin{tabular}{cccccc}
         \hline
         \hline
         Band & With LCs & Accepted LCs & Passed Test & In ATLAS \\
         \hline
         \hline
          $g$ & 35514 (65.3\%) & 34164 (96.2\%) & 70 (0.2\%) & 56 (80.0\%) \\
          $i$ & 9212 (17.0\%) & 1593 (17.3\%) & 80 (5.0\%) & 79 (98.8\%) \\
          $r$ & 30752 (56.6\%) & 29531 (96.0\%) & 165 (0.6\%) & 131 (79.4\%) \\ \hline
         \hline
    \end{tabular}
\end{table}

Once the sample based on light curve brightening was selected, we cross-matched it with the ATLAS variable star catalogue \citep{Heinze_2018} to obtain their classifications, if available. The light curves of this sample were then visually inspected, leaving only those that exhibited an overall increase in luminosity.

\begin{figure}
    \centering
    \includegraphics[width=\linewidth]{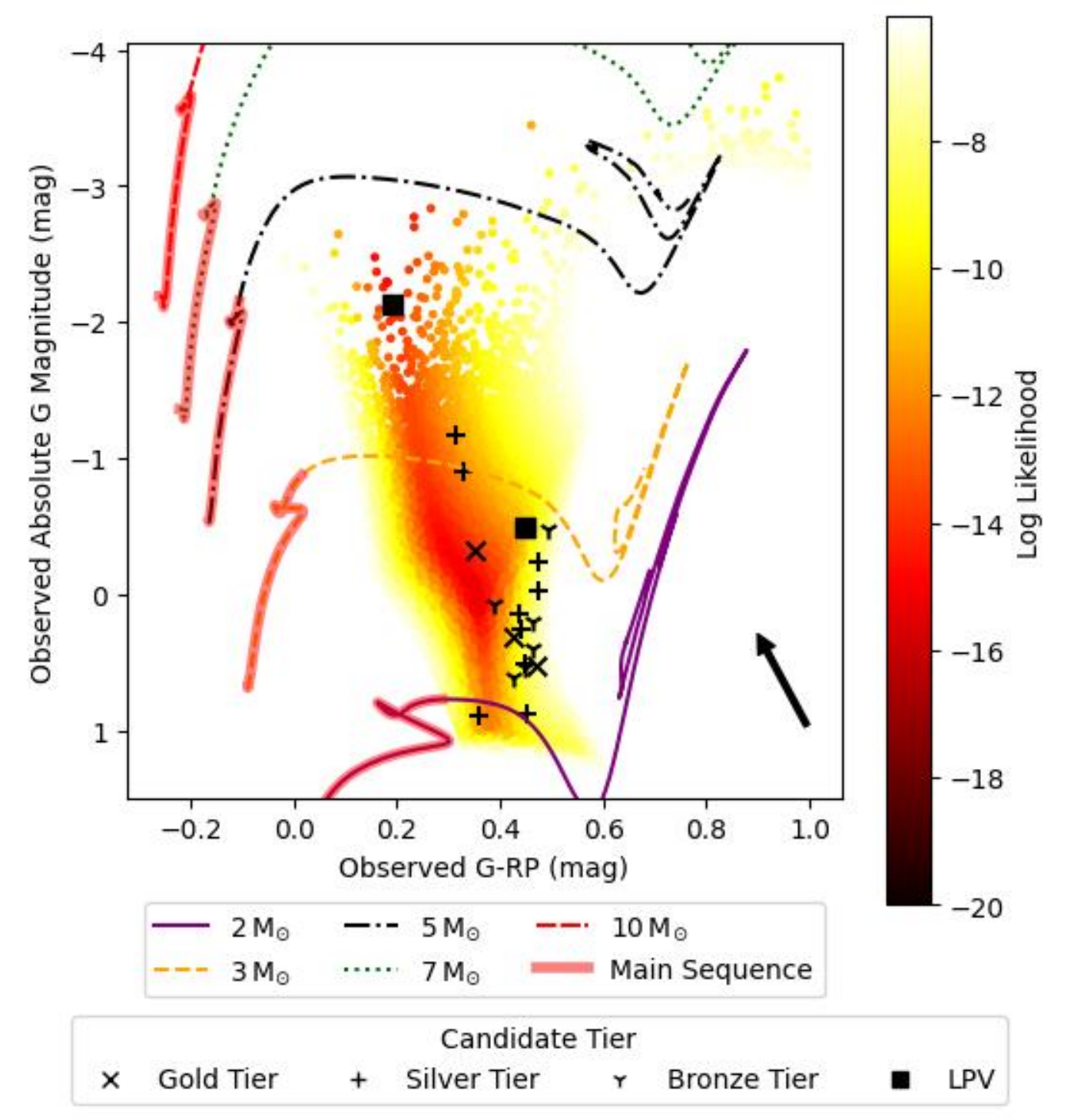}
    \caption{Observed CMD of the selected progenitor candidates with their log likelihood values of belonging to the GMM. Also plotted are MIST stellar evolution tracks \citep{MIST_0, MIST_1,paxton_2011, paxton_2013, paxton_2015, paxton_2018} from the MS to core helium burning phase for solar metallicity stars of masses $2$, $3$, $5$, $7$, and $10\,M_\odot$ with an initial/critical rotational velocity of 0.4. Represented by the arrow is the extinction and reddening correction corresponding to 1 magnitude of $V$-band extinction ($A_V$).}
    \label{fig:Candidate_Progenitors_CMD}
\end{figure}

\subsection{List of precursor candidates} \label{sec:list_candidates}

In Fig. \ref{fig:Num_Sources_Flow_Chart}, we provide a summary of the results of our LRN precursor selection strategy, where we indicate the quantity of sources and light curves at different stages of the search process. Following the method described in Section \ref{sec:Progenitor_Selection_Method}, we obtained a fitting data set from Gaia EDR3 containing approximately $9.7 \times{10^6}$ sources. From these we identified 54347 sources within the HG, shown in the CMD in Fig. \ref{fig:Candidate_Progenitors_CMD}. Single stellar evolution tracks for MS and core helium-burning phase are provided for masses between 2 and 10$\,\rm{M_{\odot}}$. The tracks correspond to MIST \citep{MIST_1} models with solar metallicity and an initial/critical rotational velocity ratio of 0.4.

In the next stage, we used the 54347 progenitor candidates as our initial sample, where we applied the methods presented in Section \ref{sec:Precursor_Selection} to identify possible precursors. Summarised in Table \ref{tab:LC numbers} are the number of light curves at the different stages of these methods. From the ZTF time-domain survey, we obtained 35514 $g$-band, 9212 $i$-band, and 30752 $r$-band light curves. From these, 34164 $g$-band (96.2 per cent), 1593 $i$-band (17.3 per cent), and 29531 $r$-band (96.0 per cent) satisfied the quality constraints described in Section \ref{sec:Light_Curve_Data} and were used for our transient identification method outlined in Section \ref{sec:Precursor_Selection_Method}. In total, 315 light curves corresponding to 294 unique sources were identified as slow-rising transients by our detection method. Of these, 247 sources were present in the ATLAS variable star catalogue, of which 9 were close binaries, 25 distant binaries, 27 irregular variables, 9 long-period variables, 136 pulsating variables, 28 multimodal pulsators, 5 sinusodials, 20 unlikely variables, and 9 stochastic variables.
Our visual inspection of the light curves of the 294 sources resulted in the identification of 21 unique sources of interest (273 false positives), further described in the next Section.

\begin{figure}
    \centering
    \includegraphics[width=\linewidth]{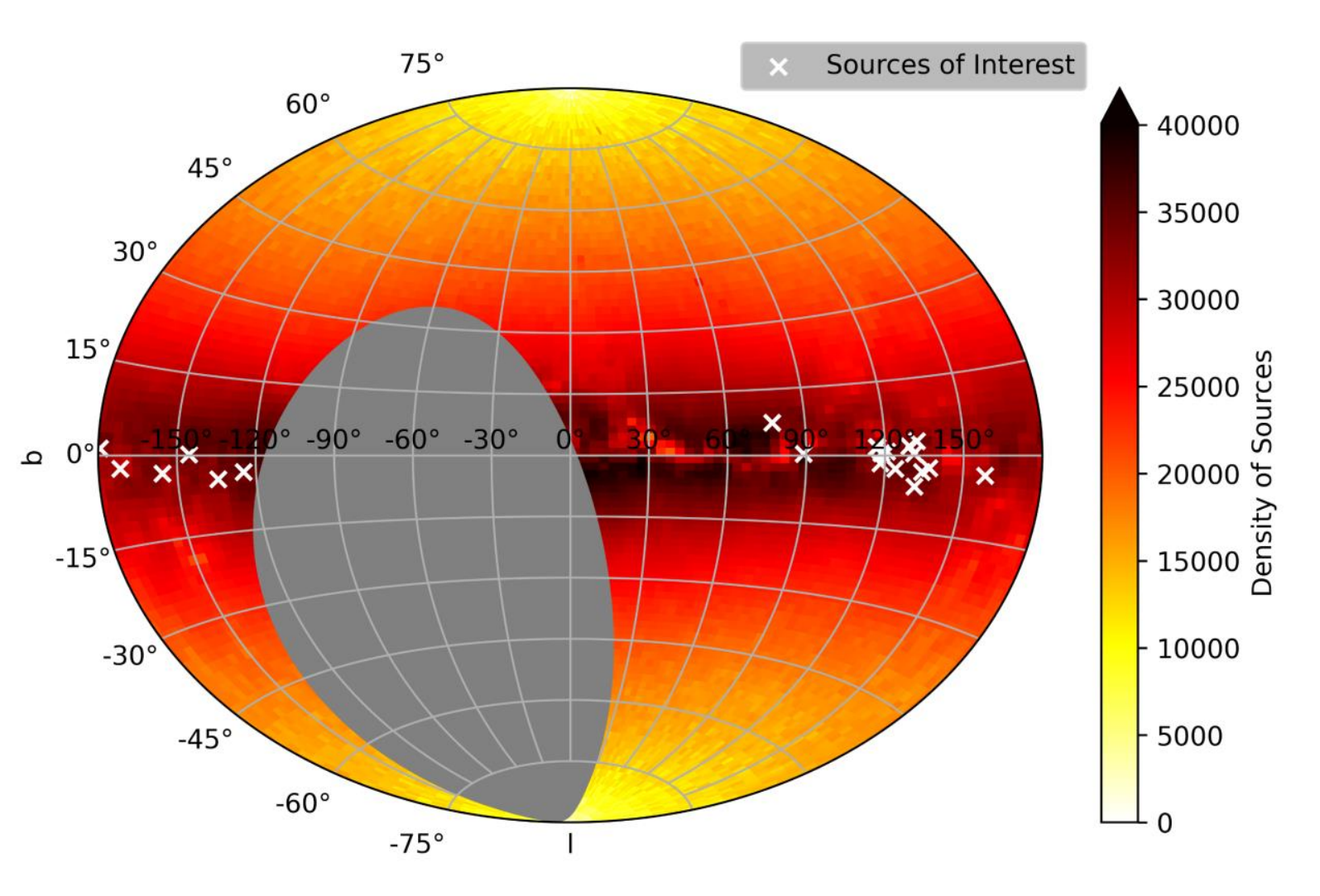}
    \caption{Sky map showing the positions of the 21 sources of interest with reference to the Galactic density of stars. The density map was constructed from the fitting data set, and the dense band represents the Galactic plane. The grey region of the sky map represents the region where no data was collected due to the declination limit of the ZTF survey.}
    \label{fig:Sources_Interest_Density_Map}
\end{figure}

\section{Archival data analysis and new observations}\label{sec:archival}

In Table \ref{tab:Interesting Sources Info}, we present data regarding our 21 sources of interest, complemented with information from existing literature available from the SIMBAD (Set of Identifications, Measurements and Bibliography for Astronomical Data) \citep[][]{Wenger_2000} database. The names of our sources of interest are constructed using their integer coordinates in degrees in the following manner: YSG\_RA\_Dec. Their distribution in the sky in Galactic coordinates is shown in Fig. \ref{fig:Sources_Interest_Density_Map}. The majority of our sources of interest lie in or near the Galactic plane, and more specifically the outer Galactic disc.

\subsection{Time-Domain Archival Data}

\begin{figure*}
    \centering
    \includegraphics[height=0.92\textheight]{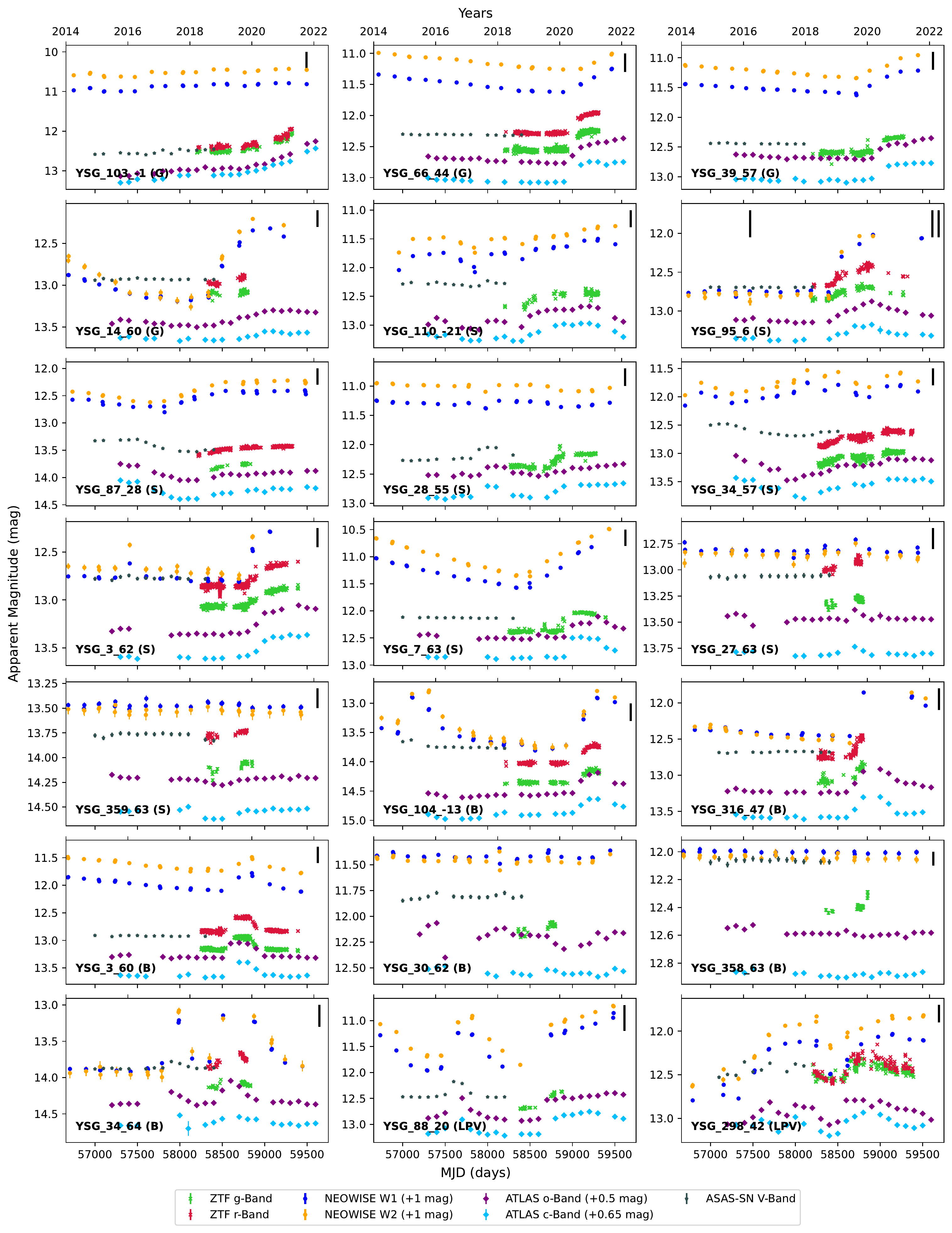}
    \caption{Optical and IR light curves of the 21 LRN precursor candidates. The ZTF \citep{Bellm_2018, Masci_2018} $g$ and $r$-band data was used in the selection of the candidates, while the ATLAS \citep{Heinze_2018,Tonry_2018,Smith_2020}, ASAS-SN \citep{Shappee_2014, Jayasinghe_2019}, and NEOWISE \citep{Mainzer_2011} data is part of our follow up investigation of these sources, and has been binned. The tier of each of the sources is denoted after the source name as G (Gold), S (Silver), B (Bronze), or LPV. Epochs of the candidate's obtained spectra are represented by the horizontal lines.}
    \label{fig:LCs}
\end{figure*}

The ZTF light curve data used in Section \ref{sec:Precursor_Selection}, was complemented with data from the optical time-domain surveys Asteroid Terrestrial-impact Last Alert System's \citep[ATLAS;][]{Heinze_2018,Tonry_2018,Smith_2020} $o$- and $c$-band forced photometry, the All-Sky Automated Survey for Supernovae \citep[ASAS-SN;][]{Shappee_2014, Jayasinghe_2019} $V$-band photometry, and the infrared Near-Earth Object Wide-Field Infrared Survey Explorer's \citep[NEOWISE;][]{Mainzer_2011} survey with $W1$ and $W2$-band photometry. This data covers a time period from 2014 up until February 2022. Fig. \ref{fig:LCs} shows the data from these surveys along with the ZTF light curves. It should be noted that the light curves have been binned with the exception of the ZTF data. For clarity, the light curves were binned using the weighted average method, where the weight of each observation is the reciprocal of its error squared. Assuming that the observations and their errors are Gaussian in nature, we combined all the measurements within the bin and assigned to each bin their joined mean and standard deviation.

\begin{figure*}
    \centering
    \includegraphics[width=\linewidth]{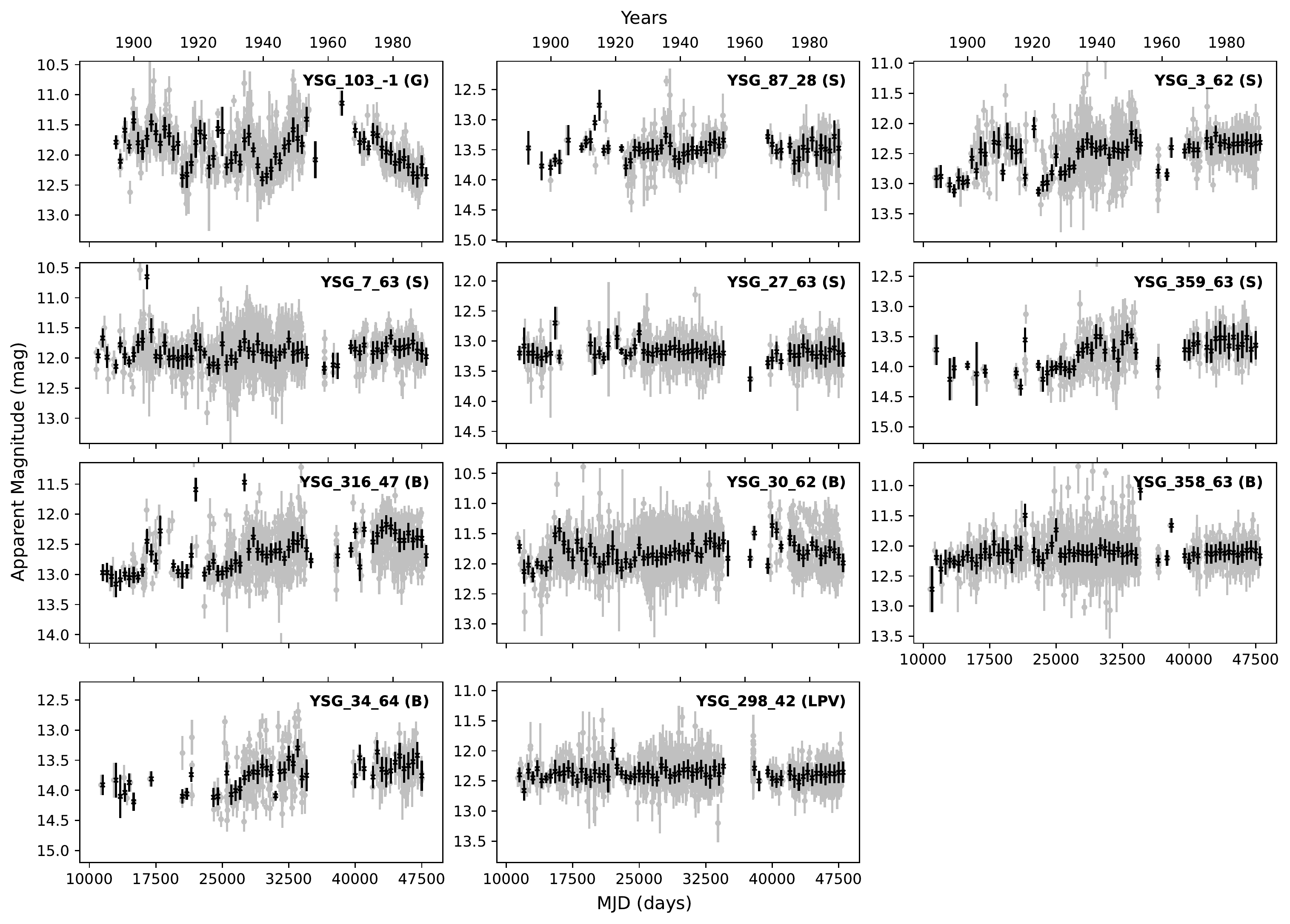}
    \caption{Available DASCH \citep{DASCH_2011} light curves of the LRN precursor candidates. The black data points are the result of binning the light curve data (grey). Furthermore, the displayed magnitude range of the light curves has been limited to $\sim\pm1.5\,$ mag about their mean magnitudes.}
    \label{fig:DASCH_LCs}
\end{figure*}

In addition, to gain insight on long time scale variability, we also searched for historical time-domain data of our precursor candidates from the Digital Access to a Sky Century @ Harvard \citep[DASCH;][]{Laycock_2010,DASCH_2011} project, covering approximately the time period from 1885 to 1992. Fig. \ref{fig:DASCH_LCs} shows the binned DASCH light curves, which were available for 11 of our 21 candidates.

\subsubsection{Periodicity search} 

In addition to a steady brightening, the LRN V1309\,Sco also exhibited the periodic signal of an eclipsing binary system. Hence, here we also conducted a periodicity search for our 21 sources of interest. In this analysis, ZTF, ATLAS, and ASAS-SN light curves were first detrended by fitting fifth order polynomials. The resulting residuals we then search for possible periodicity using the \texttt{cuvarbase} \citep{Hoffman_2017} implementations of the Lomb-Scargle periodogram \citep{Lomb_1976, Scargle_1982}, box least squares periodogram \citep{Kovacs_2002}, and conditional entropy periodogram \citep{Graham_2013} methods. We limited the range of periods searched between the period corresponding to the Nyquist frequency of the light curves and one third of the time span of the light curves. We then selected the best fitting periods as those with the highest power or lowest entropy, and which were more than 5 standard deviations from the average power or entropy of the periodograms.
From this analysis, only YSG\_103\_$-1$ and YSG\_110\_$-21$ were found to exhibit significant best fitting periods, which are 0.99 days and 43.65 days respectively.

\subsection{Candidate Classification}\label{sec:tier_system}

Each source from our sample of LRN precursor candidates was rated as Gold, Silver, or Bronze, based on their ZTF, ATLAS, NEOWISE, and ASAS-SN time-domain data. The allocated tiers represent the likelihood of each candidate being a precursor. More specifically, Gold-tiered sources are those that exhibit only an increase in brightness, Silver tiered sources are those that brighten but also experience a small decrease in brightness, and Bronze tiered sources are those that return to their pre-brightening magnitude.
Additionally, the two sources classified as LPV by ATLAS are also included for verification purposes, but these are excluded from our tier system.

\subsection{Spectral Energy Distribution}

\begin{figure*}
    \centering
    \includegraphics[width=0.99\linewidth]{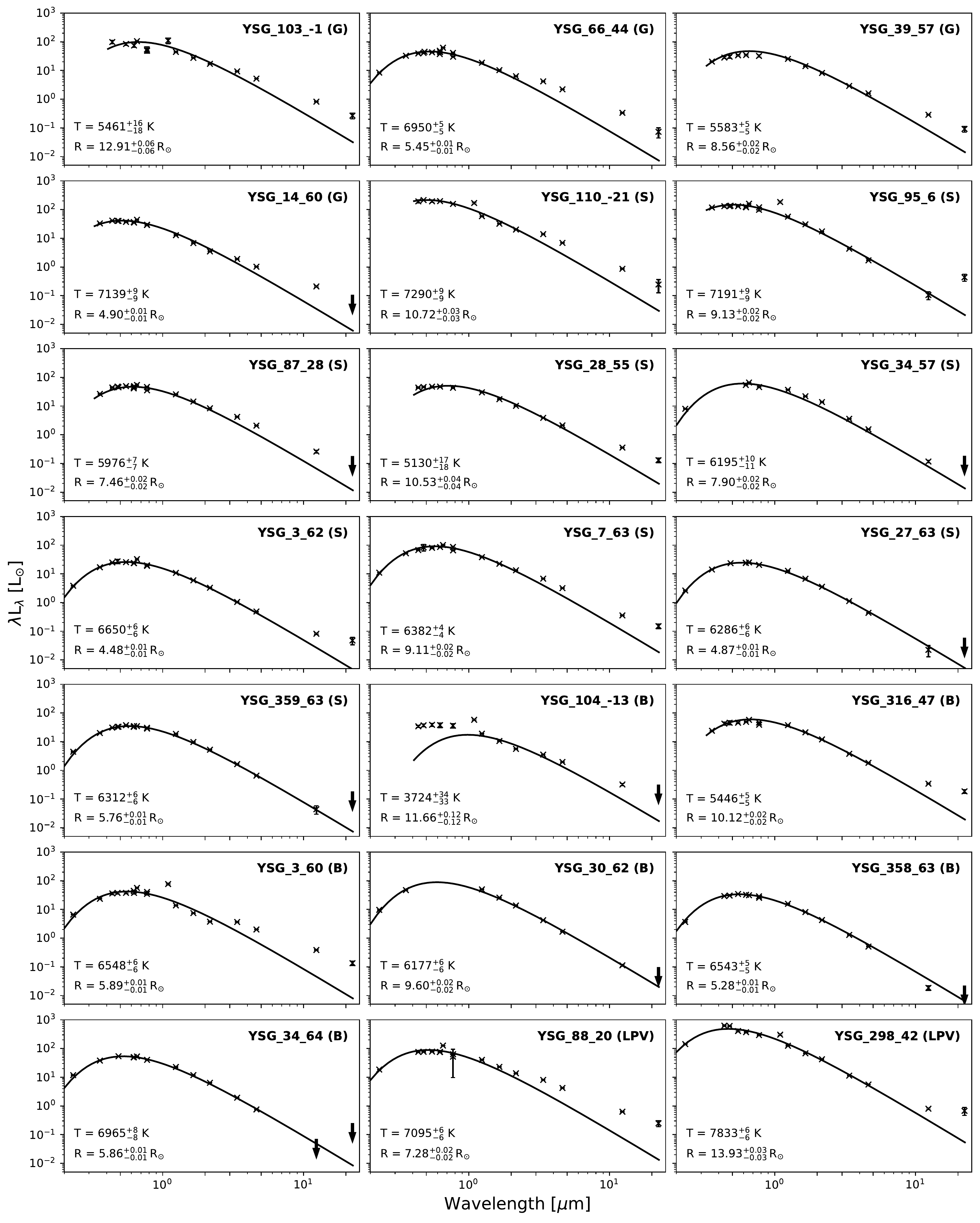}
    \caption{Spectral energy distributions (SEDs) and fitted single black-body model spectra of the 21 LRN precursor candidates. The SEDs were produced from single epoch photometry collected from APASS \citep{Henden_2015}, 2MASS \citep{Skrutskie_2006}, AllWISE \citep{Wright_2010, Mainzer_2011}, GALEX \citep{Bianchi_2011}, IPHAS \citep{Drew_2005}, and UVEX \citep{Groot_2009}. Observations representing upper luminosity limits are denoted with downwards arrows.}
    \label{fig:SEDs}
\end{figure*}

Spectral energy distributions (SEDs) were constructed for each of our precursor candidates. We made use of the American Association of Variable Star Observers Photometric All-Sky Survey DR9 and DR10's \citep[APASS;][]{Henden_2015} $B$, $V$, $g'$, $r'$, and $i'$-band photometry, the Two Micron All Sky Survey's \citep[2MASS;][]{Skrutskie_2006} $J$, $H$, and $K_{\rm{s}}$ photometry, AllWISE's \citep{Wright_2010, Mainzer_2011} $W1$, $W2$, $W3$, and $W4$-band photometry, the Galaxy Evolution Explorer's \citep[GALEX;][]{Bianchi_2011} $NUV$-band photometry, the Isaac Newton Telescope Photometric H$\alpha$ Survey's \citep[IPHAS;][]{Drew_2005} $i$-, H$\alpha$, and $r$-band photometry, and the UV-Excess Survey's \citep[UVEX;][]{Groot_2009} $g$-, $r$-, and RGO $U$-band photometry.
To construct the SEDs from the photometric data, we converted the magnitude measurements to fluxes using the Gaia EDR3 distances. The resulting SEDs are shown in Fig. \ref{fig:SEDs}. It should be noted that these sources are variable in nature and the SEDs were constructed from several different epochs ranging from 1997 to 2021. Therefore, the given SEDs do not fully represent the sources at any given moment.

To investigate possible deviations from a single emission component for our candidates, we fitted single black-body spectra to the constructed SEDs. We made use of the \texttt{Python} package \texttt{Utils} \citep{Utils_Nblago}, which applies a Markov chain Monte Carlo approach to obtain the best fit temperature and radius of the black body emission. In this analysis we did not consider the effects of extinction and we did not exclude any observations where an excess in infrared (IR) was noted. The best fits and values are included with the observational SED data in Fig. \ref{fig:SEDs}, with the best fit values also displayed in Table \ref{tab:SED_best_fits}. It should be noted that the errors associated with the best fit temperatures and radii are purely statistical in nature and do not account for any systematic uncertainties.

\newpage
\renewcommand{\tabcolsep}{0.05cm}
\def\arraystretch{1.7}
\begin{landscape}
\begin{table}
    \caption{Information of the sources of interest collected from Gaia EDR3 \citep{Gaia_EDR3_2021}, our existing literature search, and follow up investigations. The source of the data is \citet{Gaia_EDR3_2021} unless stated otherwise. The V-band extinction values, $A_V$, were calculated as $A_V = 3.1 \times E(B-V)$, where $E(B-V)$ are the line-of-sight reddening provided by \citet{Schlafly_2011}.\\
    H$\alpha$ Em = H$\alpha$ Emission, H$\alpha$ Ab = H$\alpha$ Absorption, X-ray = X-ray emission, LPV = Long Period Variable, Em star = Emission-Line Star, Infra = Infrared Source, IRR = Irregular variable, Dubious = Probably not a variable, VAR = Variable Star.\\
    (1) \citep{Bailer_Jones_2021} , (2) \citet{Schlafly_2011} , 
    (3) \citet{Robertson_1989}, (4) \citet{pavlinsky2021srgartxc}, 
    (5) \citet{Evans_2020}, (6) \citep{LAMOST_2015}, 
    (7) \citet{Kohoutek_1999}, (8) \citet{MacConnell_1983}, 
    (9) \citet{Coyne_1978}, (10) \citet{Stephenson_1977}, 
    (11) \citet{Gonzalez_1956_59_90}, (12) \citet{Coyne_1983}, 
    (13) \citet{Gonzalez_1956_103_180}, (14) \citet{Wenger_2000}, 
    (15) \citet{Heinze_2018}, (16) \citet{Gaia_DR3}, (SP) Follow-up spectra \\
    }
    \label{tab:Interesting Sources Info}
    \fontsize{7.4pt}{7.4pt}\selectfont{}
    \begin{tabular}{lcrccccccccccccccccccc}
    
         \hline
         \hline
         Source & Gaia source id & RA & Dec & Distance (1) & PM$_{\rm{RA}}$ & PM$_{\rm{Dec}}$ & $G$ & $G-RP$ & $A_g$ (16) & $A_V$ (2) & Properties & Simbad Class. (14) & ATLAS Class. (15) & Tier \\
         & & ($^{\circ}$) & ($^{\circ}$) & (pc) & (mas/yr) & (mas/yr) & (mag) & (mag) & (mag) & (mag) & & (Secondary Class.) & & \\
         \hline
         \hline
         
         YSG\_103\_$-$1 & 3112116539430781824 & 103.80181 & $-1.4815$ & $3619_{-215}^{+283}$ & $-0.62\,\pm\,$0.02 & 0.77$\,\pm\,$0.02 & 12.47 & 0.35 & $0.6072_{-0.0029}^{+0.0054}$ & 1.86 & H$\alpha$ Em (3, SP), X-ray (4, 5) & LPV (Em Star, Infra) & LPV & Gold\\

         YSG\_66\_44 & 253403143486587136 & 66.59266 & 44.0087 & $2391_{-82}^{+113}$ & $-0.64\,\pm\,$0.02 & $-1.07\,\pm\,$0.01 & 12.21 & 0.43 & $1.9979_{-0.0032}^{+0.0028}$ & 1.89 & H$\alpha$ Em (6, 7, 13, SP) & Em Star (Infra) & -- & Gold\\

         YSG\_39\_57 & 457906018421337344 & 39.56234 & 57.4849 & $2162_{-50}^{+42}$ & $-1.14\,\pm\,$0.01 & 0.41$\,\pm\,$0.01 & 12.20 & 0.47 & -- & 2.51 & H$\alpha$ Em (7, 12, SP) & Em Star (Infra) & -- & Gold\\

         YSG\_14\_60 & 426462008788804096 & 14.92932 & 60.1786 & $3053_{-127}^{+157}$ & $-0.93\,\pm\,$0.01 & $-1.40\,\pm\,$0.01 & 12.89 & 0.28 & -- & 1.33 & H$\alpha$ Em (SP) & -- & -- & Gold\\
         
         YSG\_110\_$-$21 & 2929639837342247296 & 110.31019 & $-21.3153$ & $5272_{-369}^{+341}$ & $-0.51\,\pm\,$0.01 & 1.57$\,\pm\,$0.01 & 12.43 & 0.31 & $1.6653_{-0.0045}^{+0.0018}$ & 1.89 & H$\alpha$ Em (SP) & Star (Infra) & IRR & Silver\\

         YSG\_95\_6 & 3323902472007123840 & 95.98428 & 6.2353 & $5032_{-415}^{+505}$ & $0.01\,\pm\,$0.02 & 0.03$\,\pm\,$0.02 & 12.61 & 0.33 & $1.4331_{-0.0025}^{+0.0029}$ & 1.61 & H$\alpha$ Ab (6, SP), H$\alpha$ Em (SP) & -- & -- & Silver\\
         
         YSG\_87\_28 & 3443268615416009472 & 87.82820 & 28.9491 & $4319_{-419}^{+486}$ & 0.3$\,\pm\,$0.02 & $-1.01\,\pm\,$0.01 & 13.43 & 0.44 & $1.7519_{-0.0025}^{+0.0023}$ & 1.98 & H$\alpha$ Em (6, SP) & -- & Dubious & Silver\\
         
         YSG\_28\_55 & 504717657256222080 & 28.47645 & 55.7053 & $2412_{-87}^{+85}$ & $-0.70\,\pm\,$0.01 & $-1.29\,\pm\,$0.02 & 12.06 & 0.44 & $1.8834_{-0.0049}^{+0.0016}$ & 1.09 & H$\alpha$ Em (7, 8, 9, 10, SP) & Em Star (Infra) & Dubious & Silver\\

         YSG\_34\_57 & 458686053202171520 & 34.35626 & 57.7478 & $3531_{-192}^{+176}$  & $-1.06\,\pm\,$0.01 & $-0.07\,\pm\,$0.01 & 12.72 & 0.48 & $2.1314_{-0.0049}^{+0.0045}$ & 1.92 & H$\alpha$ Em (SP) & -- & -- & Silver\\
         
         YSG\_3\_62 & 431333223256033536 & 3.11473 & 62.9781 & $2412_{-67}^{+86}$  & $-1.91\,\pm\,$0.01 & $-0.96\,\pm\,$0.01 & 12.80 & 0.36 & -- & 2.48 & H$\alpha$ Em (7, 11, SP) & Em Star (Infra) & -- & Silver\\

         YSG\_7\_63 & 430894586836432768 & 7.99930 & 63.5143 & $2781_{-78}^{+92}$  & $-2.90\,\pm\,$0.01 & $-0.75\,\pm\,$0.01 & 11.99 & 0.47 & -- & 4.03 & H$\alpha$ Em (7, SP) & Em Star (Infra) & -- & Silver\\

         YSG\_27\_63 & 512160491980348928 & 27.39134 & 63.9136 & $2626_{-105}^{+133}$  & $-0.95\,\pm\,$0.01 & $-0.17\,\pm\,$0.01 & 12.97 & 0.45 & $1.7019_{-0.0022}^{+0.0050}$ & 4.19 & H$\alpha$ Ab (SP) & -- & -- & Silver\\
         
         YSG\_359\_63 & 2016198768498312192 & 359.73548 & 63.9497 & $4364_{-241}^{+283}$ & $-2.07\,\pm\,$0.01 & $-1.11\,\pm\,$0.01 & 13.71 & 0.45 & $1.6029_{-0.0017}^{+0.0018}$ & 2.82 & H$\alpha$ Ab (SP) & -- & -- & Silver\\
         
         YSG\_104\_$-$13 & 2949520725556164864 & 104.46141 & $-13.2202$ & $5174_{-390}^{+383}$ & $-0.59\,\pm\,$0.02 & 1.41$\,\pm\,$0.02 & 13.97 & 0.46 & -- & 2.02 & H$\alpha$ Em (SP) & Star & IRR & Bronze\\
         
         YSG\_316\_47 & 2165231827256902144 & 316.54535 & 47.7675 & $3203_{-94}^{+96}$ & $-4.01\,\pm\,$0.01 & $-3.56\,\pm\,$0.01 & 12.73 & 0.46 & $1.7959_{-0.0005}^{+0.0005}$ & 7.29 & H$\alpha$ Em (SP) & -- & -- & Bronze\\
         
         YSG\_3\_60 & 429230171057310720 & 3.09651 & 60.8852 & $2826_{-109}^{+90}$ & $-2.78\,\pm\,$0.01 & $-1.18\,\pm\,$0.01 & 12.79 & 0.45 & $0.0211_{-0.0127}^{+0.0147}$ & 3.01 &  H$\alpha$ Em (SP) & -- & -- & Bronze\\
         
         YSG\_30\_62 & 508677720181849984 & 30.43455 & 62.0812 & $2745_{-145}^{+159}$ & 0.37$\,\pm\,$0.01 & $-1.00\,\pm\,$0.02 & 11.72 & 0.49 & -- & 3.72 & & Star (Infra) & -- & Bronze \\
         
         YSG\_358\_63 & 2016194056905090688 & 358.91366 & 63.8671 & $1956_{-34}^{+36}$ & $-4.51\,\pm\,$0.01 & $-1.33\,\pm\,$0.01 & 12.06 & 0.43 & $1.6536_{-0.0047}^{+0.0038}$ & 3.50 & H$\alpha$ Ab (SP) & Star (Infra) & -- & Bronze\\
        
         YSG\_34\_64 & 515178445243685248 & 34.45508 & 64.0326 & $5694_{-418}^{+653}$ & $-1.46\,\pm\,$0.01 & 0.50$\,\pm\,$0.01 & 13.85 & 0.39 & $1.6174_{-0.0024}^{+0.0025}$ & 2.42 & H$\alpha$ Em (SP) & -- & -- & Bronze\\
         
         YSG\_88\_20 & 3423114086241636992 & 88.97008 & 20.7680 & $3559_{-243}^{+346}$ & $-0.32\,\pm\,$0.02 & $-1.34\,\pm\,$0.02 & 12.27 & 0.45 & $1.9202_{-0.0068}^{+0.0058}$ & 2.76 & H$\alpha$ Em (7, 13, SP) & Em Star & LPV & LPV\\
         
         YSG\_298\_42 & 2075547515637225344 & 298.78552 & 42.6502 & $7968_{-1064}^{+1293}$  & $-3.23\,\pm\,$0.02 & $-5.09\,\pm\,$0.02 & 12.38 & 0.19 & $1.1047_{-0.0030}^{+0.0037}$ & 1.09 & H$\alpha$ Em (SP) & VAR (Infra) & LPV & LPV \\
         \hline
         \hline
         
    \end{tabular}
\end{table}
\end{landscape}

\def\arraystretch{1.2}
\begin{table}
    \centering
    \caption{Resulting best fit temperatures and radii corresponding to the best fit single black-body spectra fitted to the observational SEDs. The provided errors do not account for systematic uncertainties and are purely statistical in nature.}
    \begin{tabular}{lcc}
         \hline
         \hline 
         
          Source & Temperature (K) & Radius (\Rsun{})\\
          
          \hline
          \hline 
         
          YSG\_103\_$-$1 & $5461_{-18}^{+16}$ & $12.91 \pm 0.06$ \\
          YSG\_66\_44 & $6950\pm5$ & $5.45 \pm 0.01$ \\
          YSG\_39\_57 & $5583\pm5$ & $8.56\pm0.02$ \\
          YSG\_14\_60 & $7139\pm9$ & $4.90\pm0.01$ \\
          YSG\_110\_$-$21 & $7290\pm9$ & $10.72\pm0.03$ \\
          YSG\_95\_6 & $7191\pm9$ & $9.13\pm0.02$ \\
          YSG\_87\_28 & $5976\pm7$ & $7.46\pm0.02$ \\ 
          YSG\_28\_55 & $5130_{-18}^{+17}$ & $10.53\pm0.04$ \\
          YSG\_34\_57 & $6195_{-11}^{+10}$ & $7.90\pm0.02$ \\
          YSG\_3\_62 & $6650\pm 6$ & $4.48\pm0.01$ \\
          YSG\_7\_63 & $6382\pm4$ & $9.11\pm0.02$ \\
          YSG\_27\_63 & $6286\pm6$ & $4.87\pm0.01$ \\
          YSG\_359\_63 & $6312\pm6$ & $5.76\pm0.01$ \\
          YSG\_104\_$-$13 & $3724_{-33}^{+34}$ & $11.66\pm0.12$ \\
          YSG\_316\_47 & $5446\pm5$ & $10.12\pm0.02$ \\
          YSG\_3\_60 & $6548\pm6$ & $5.89\pm0.01$ \\
          YSG\_30\_62 & $6177\pm6$ & $9.60\pm0.02$ \\
          YSG\_358\_63 & $6543\pm5$ & $5.28\pm0.01$ \\
          YSG\_34\_64 & $6965\pm8$ & $5.86\pm0.01$ \\
          YSG\_88\_20 & $7095\pm6$ & $7.28\pm0.02$ \\
          YSG\_298\_42 & $7833\pm6$ & $13.93\pm0.03$ \\
          
          \hline
          \hline 
         
    \end{tabular}
    
    \label{tab:SED_best_fits}
\end{table}

\subsection{Spectroscopic follow-up} \label{sec:followup}

Low-resolution spectra was obtained for all Golden and Silver candidates in our sample, and approximately half of the Bronze and LPV candidates. Data on northern candidates was obtained with the Spectrograph for the Rapid Acquisition of Transients instrument \citep[SPRAT;][]{SPRAT_2014SPIE} at the Liverpool Telescope in La Palma \citep[LT;][]{LT_2004SPIE}. Data on southern targets were obtained with the Mookodi instrument at the 1\,m Lesedi telescope (which is a SPRAT-based low-resolution spectrograph), and with the low-resolution Spectrograph Upgrade: Newly Improved Cassegrain \citep[SpUpNic;][]{SpUpNic} mounted on the 1.9\,m telescope, both located at the South African Astronomical Observatory (SAAO).  The instrumental resolution of Mookodi data is of $\simeq$875\,\kms, SpUpNic (Grating 7) of 250\,\kms, and of SPRAT on LT of $\simeq$820\,\kms. The observation date and instrument for each spectrum is shown in Table \ref{tab:Spectra_Info}.

\def\arraystretch{1}
\begin{table}
    \caption{Observation date and instrument of the spectra obtained for the precursor candidates in the order they appear in Fig. \ref{fig:spectra}.}
    \label{tab:Spectra_Info}
    \centering
    \begin{tabular}{llc}
         \hline
         \hline
         Source & Date & Instrument \\
         \hline
         \hline 
         
         YSG\_103\_$-$1 & 2021-10-06T03:07:51 & SpUpNic \\
         YSG\_66\_44 & 2022-02-11T20:14:37 & SPRAT \\
         YSG\_39\_57 & 2022-02-10T20:01:24 & SPRAT \\
         YSG\_14\_60 & 2022-02-11T21:10:54 & SPRAT \\
         YSG\_110\_$-$21 & 2022-04-17T18:04:14 & SpUpNic \\
         YSG\_95\_6 & 2016-03-21T00:00:00 & LAMOST \\
         YSG\_95\_6 & 2022-02-04T00:13:06 & Mookodi \\
         YSG\_95\_6 & 2022-04-17T18:33:34 & SpUpNic \\
         YSG\_87\_28 & 2022-02-12T20:41:19 & SPRAT \\
         YSG\_28\_55 &  2022-02-11T21:20:03 & SPRAT \\
         YSG\_34\_57 &  2022-02-11T20:45:46 & SPRAT \\
         YSG\_3\_62 & 2022-02-12T20:17:13 & SPRAT \\
         YSG\_7\_63 & 2022-02-12T19:54:27 & SPRAT \\
         YSG\_27\_63 & 2022-02-11T20:36:35 & SPRAT \\
         YSG\_359\_63 & 2022-02-13T19:47:36 & SPRAT \\ 
         YSG\_104\_$-$13 & 2022-04-17T19:07:41 & SpUpNic \\
         YSG\_316\_47 & 2022-04-24T03:42:57 & SPRAT \\
         YSG\_3\_60 &  2022-02-15T19:43:15 & SPRAT \\
         YSG\_358\_63 & 2022-02-13T19:54:25 & SPRAT \\
         YSG\_34\_64 & 2022-03-05T20:10:29 & SPRAT \\
         YSG\_88\_20 & 2022-02-04T00:08:21 & Mookodi \\
         YSG\_298\_42 & 2022-04-24T03:33:45 & SPRAT \\
        
         \hline
         \hline
    \end{tabular}
\end{table}

\begin{figure*}
    \centering
    \includegraphics[width=\linewidth]{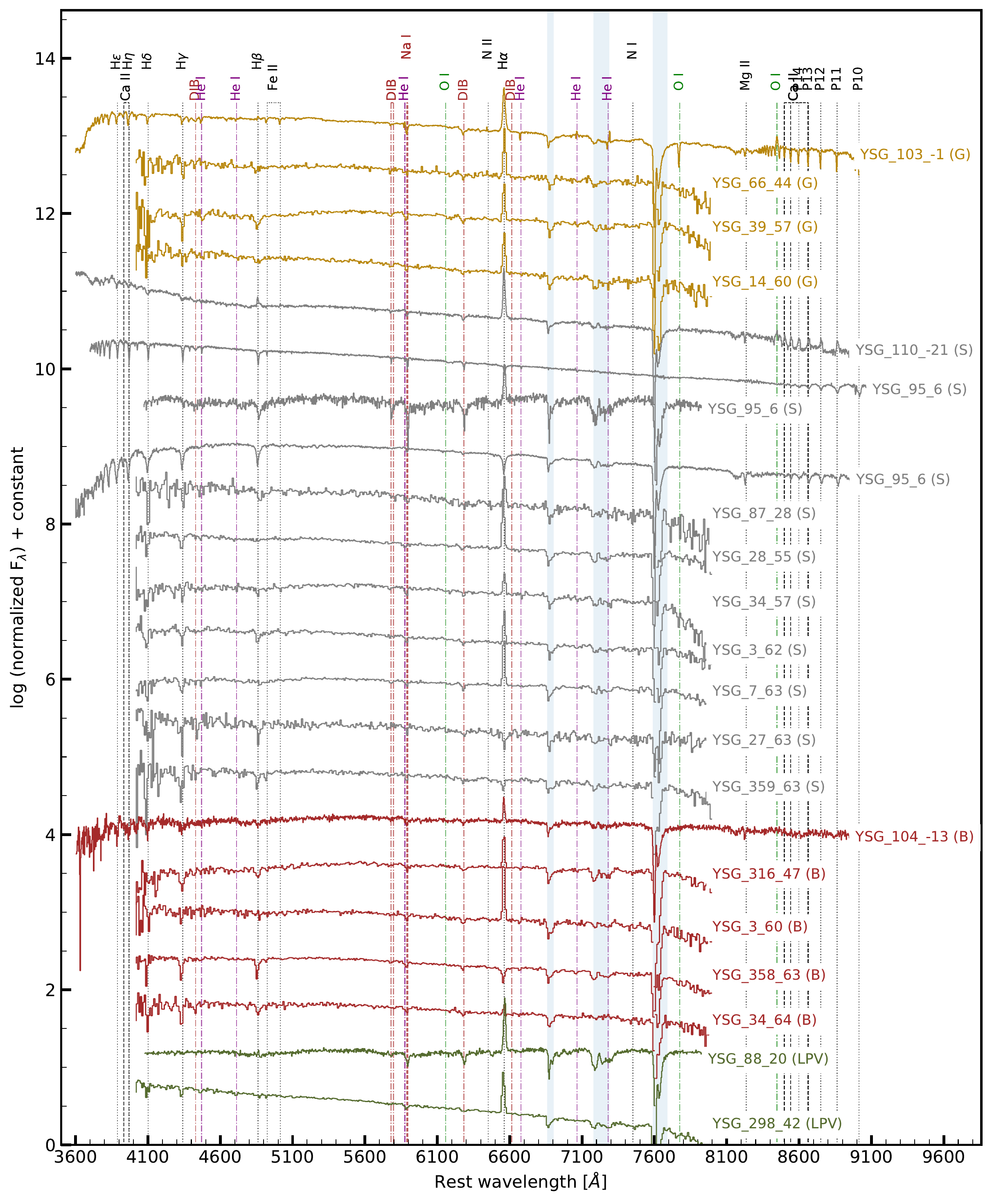}
    \caption{Optical follow-up spectra for the precursor candidates. The main emission and absorption lines are indicated. The areas with strong telluric absorption is shown by the shaded rectangles. The candidates are colour-coded by tier, which is also indicated in parenthesis. The source YSG\_95\_6 was observed in three different epochs, and the spectra are displayed in chronological order.}
    \label{fig:spectra}
\end{figure*}

\begin{figure*}
    \centering
    \includegraphics[width=\linewidth]{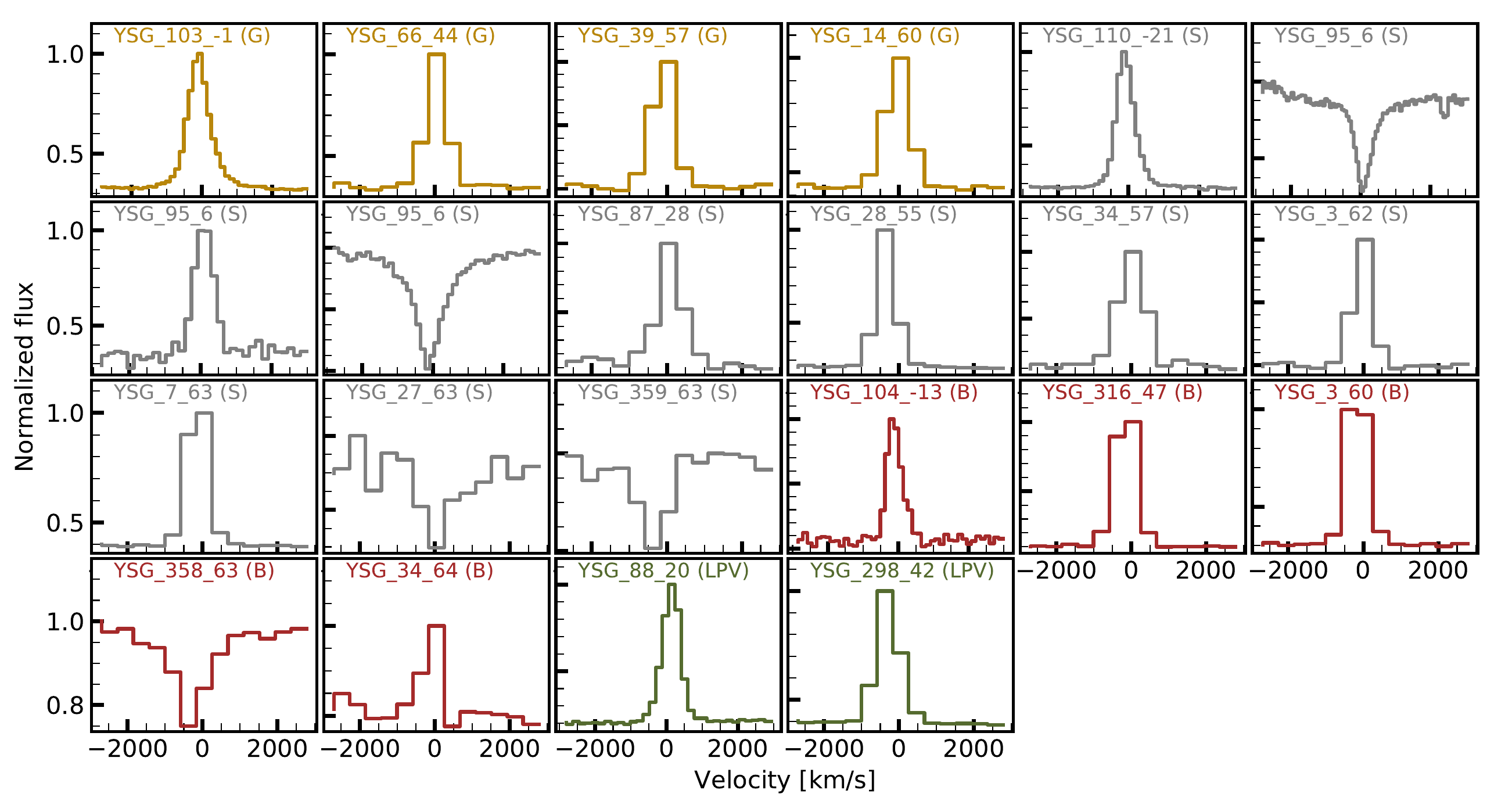}
    \caption{Flux normalized H$\alpha$ velocity profile for the observed candidate sources. All but three sources show H$\alpha$ in emission. Due to the low resolution of the spectra, the widths of the profiles reproduce the instrumental resolution of the spectrographs. In some cases, the centre of the line presents an unusually large shift from zero velocity, likely caused by inaccuracies with the wavelength calibration. }
    \label{fig:halpha}
\end{figure*}

SpUpNic and Mookodi data were reduced using a custom developed data reduction pipeline in \texttt{Python}. SPRAT data were reduced using the automated LT pipeline \citep{LT_datared}. The sequence of all spectra is displayed in Fig. \ref{fig:spectra}, and the velocity of the normalized flux of the H$\alpha$ profile is shown in Fig. \ref{fig:halpha}. Although the spectra were corrected for heliocentric velocity, some sources still show large shifts ($\sim-$500\,\kms) in the rest-frame velocity of this line. This shift should be interpreted with caution due to the low resolution of the spectra and the likely possibility that it is caused by inaccurate wavelength calibration.

\section{Results}\label{sec:Results}

\subsection{Characteristics for the sample of precursor candidates}

\subsubsection{Gold tiered candidates}

The optical light curves of our Gold tiered candidates in Fig. \ref{fig:LCs} show an increase in brightness of a few tenths of a magnitude within the last 2 to 7 years. Older data indicates a quiescent behaviour in the optical bands. Interestingly, the IR NEOWISE light curves are more varied in nature and show that 3 of the 4 sources faded slowly before the optical outburst, but then the optical increase was also accompanied by IR emission.
Our periodicity searches of these light curves resulted in the identification of a possible 1 day period for YSG\_103\_$-$1. However, it should be noted that this period was identified from the ground-based ZTF data, where a 1 day period is a very common alias. 

Historical data for the candidate YSG\_103\_$-$1 was also found in the DASCH archive. Its light curve, shown in Fig. \ref{fig:DASCH_LCs}, shows an irregular variability on timescales of decades. For example from $\sim$1940 to 1950 the source appears to brighten from 12.5\,mag to 11.5\,mag, but then it fades back to 12.5 in about 30 years. In fact, this star is also known as V520\,Mon, and it was already identified as variable several decades ago \citep{Kukarkin1968IBVS,Kukarkin1971GCVS3}. 

The UV, optical, and IR SEDs for our Gold tiered candidates, shown in Fig. \ref{fig:SEDs}, appear to have a good match to a single black-body emission-model in the optical wavelengths. The only exception is YSG\_103\_$-$1, which has a blue flux-excess for wavelengths shorter than 7000\,\AA. This source was also found to be a variable soft X-ray emitter in the Swift Point-Source catalogue \citep[2SXPS;][]{Evans_2020}. At longer wavelengths, all gold candidates show a clear excess redward of $\gtrsim1.8\rm{\mu m}$, which indicates the existence of an additional emission IR component, likely associated with a disk containing warm dust. 

Our spectroscopic follow up results, presented in Fig. \ref{fig:spectra}, show that all four Gold candidates are H$\alpha$ emitters, three of which had previously been identified by photometric H$\alpha$ surveys (see Table \ref{tab:Interesting Sources Info}). In addition, YSG\_103\_$-$1 also shows emission in H$\beta$ and a mixture of absorption and emission lines related to \io{He}{i}. All four spectra show absorption related to diffuse interstellar bands (DIBs) and Na, which indicates a moderate amount of extinction is present.

\subsubsection{Silver Tiered Candidates}

By definition (see Section \ref{sec:tier_system}), the light curves for our Silver candidates have experienced an increase in brightness, but since then they have either entered a plateau, or initiated a dimming phase, as shown in Fig. \ref{fig:LCs}. In general, the optical brightening of the light curves is less than $\sim0.5$mag over $\sim$1-2 years. Longer baseline ASAS-SN and ATLAS data show that these sources had other non-periodic variability episodes in the previous 4\,years that were not covered by ZTF data. Except for the two candidates YSG\_27\_63 and YSG\_359\_63, the mid-IR NEOWISE data also shows larger variations that are correlated with the recent optical activity. Our periodicity search of these optical light curves yielded a single significant period for candidate YSG\_110\_$-21$ of 43.65 days.

Of the 9 Silver candidates 5 have available DASCH light curves, which are presented in Fig. \ref{fig:DASCH_LCs}. YSG\_87\_28 appears to exhibit an irregular variability, while the other 4 candidates experience only small fluctuations to their brightness trends. 

Similar to the Gold sample, 7 out of 9 Silver candidates also showed an IR excess emission component. The two exceptions are YSG\_27\_63, and YSG\_359\_63, which show a good agreement with a single emission component. Our spectroscopic follow-up shows that these two sources also lack the H$\alpha$ emission detected for all the other Silver candidates.

All Silver candidates are seen to be H$\alpha$ emitters from our follow-up spectra, shown in Fig. \ref{fig:spectra}, with the exception of YSG\_27\_63, and YSG\_359\_63. The candidate YSG\_95\_6 shows variable emission. In 2016 it was observed as part of the LAMOST survey \citep{LAMOST_2015}, and was classified as a B9 type star with H$\alpha$ absorption. Our follow-up spectra taken on 2022 Feb 04 and on 2022 Mar 21 show that the source initially developed H$\alpha$ emission, which disappeared in the later spectrum. The light curve of this candidate in Fig. \ref{fig:LCs} shows that the spectra were taken before, during, and after its outburst, which took place in 2017.

\subsubsection{Bronze Tiered Candidates}

The optical light curves of our six Bronze candidates initially brighten by a few tens of magnitude, but then return to the quiescent state. However, most of them still show enhanced emission in mid-IR wavelengths, as the optical light is likely reprocessed on longer timescales by the dusty stellar environment. Previous episodes of enhanced activity are detected for YSG\_104\_$-13$ and YSG\_30\_62. In addition the IR light curve of YSG\_34\_64 shows $\sim$1\,mag variability in between consecutive epochs of NEOWISE data, taken $\sim$6\,months apart. Our periodicity searches of the optical light curves returned no periods of significance.

There are four bronze candidates with available DASCH light curves, presented in Fig. \ref{fig:DASCH_LCs}, which also show possible long-time variations over the $\sim 100$ year span. In any case, the variability in the binned data is always below 1\,mag.

The SEDs and black-body fits of our Bronze candidates show that half of the sources have excess at IR wavelengths. In addition, there is also an excess at shorter wavelengths for the candidate YSG\_104\_$-$13. For two candidates, the model seems to predict more flux than provided by the observational SED. This can be attributed to the different epochs of the surveys the data was obtained from.

Optical low-resolution spectroscopy (see Fig. \ref{fig:spectra}) was obtained for three of the six Bronze candidates. We detected absorption in H$\beta$, and in lines identified with DIB and Na, likely caused by extinction. Two of them, YSG\_104\_$-$13 and YSG\_34\_64, show emission in H$\alpha$. In the case of YSG\_34\_64, this emission line is not associated with IR excess.

\subsubsection{Non-Tiered Candidates}

Two of our candidates were not assigned to a tier, as they were classified as LPVs in the ATLAS variable star catalogue. However, in this section we re-examine this classification. The optical and IR light curves for YSG\_88\_20 and YSG\_298\_42, are shown in Fig. \ref{fig:LCs}. Both exhibit variations, but these do not appear to be periodic. For example, the second outburst for YSG\_88\_20 is still active after three years in our data, while the previous one returned to quiescence in just two years. 
The light curves of YSG\_298\_42 also exhibit non-periodic variability in the optical. In addition, before 2016, its IR emission was about 0.75\,mag fainter than before its last active episode starting in 2017. For this last source, the DASCH light curve shows that star's magnitude trend was fairly constant between 1890$-$1990, displaying no signs of longer term periodicity. 

SEDs of the two LPVs are presented in Fig. \ref{fig:SEDs}, where both display show IR excess as compared with the fitted single black-body models, and YSG\_298\_42 also appears to have a stronger flux at shorter wavelengths. Our spectroscopic follow-up shows that both candidates exhibit H$\alpha$ in emission.

\section{Discussion}\label{sec:Discussion}

\subsection{Nature of the candidates}
This is the first study to search for LRN precursors among candidate YSG progenitor sources. Although our selection process was only based on the position in the HR diagram and the variability of the sources, our results show that many of our candidates share similar characteristics, hinting at a possibly similar origin, which we discuss below.

\subsubsection{Be stars}
From our spectroscopic follow-up sample of 20 objects, 16 are H$\alpha$ emitters, and one also shows emission in one of the epochs taken during the activity episode. Although the resolution of our spectra does not allow for careful identification of narrow lines, the strong absorption present for Balmer lines other than H$\alpha$ (and occasionally H$\beta$) indicates hotter A-type, or even B-type (where \io{He}{i} lines are also detected) stars. The observation of strong absorption related to DIBs and Na in our follow-up spectra suggests that several of our candidates have a significant amount of line-of-sight extinction, and therefore appear in the HG. In this case, these candidates would correspond to reddened Be and A shell stars, which are rapidly rotating stars that have formed a disk on their equatorial plane, likely through pulsation or by radiatively-driven winds \citep{Porter2003PASP}. The hot circumstellar gas in this disk is the main cause for the emission lines, which are variable in nature, and the IR excess is caused by free$-$free and free$-$bound emission from the gas in the disk \citep{Gehrz1974ApJ}. In some cases, considerably larger IR excesses were alternatively explained by the existence of a cold companion, or the presence of a circumstellar dusty disk \citep{Dougherty1994AA}.

Among possible Be candidates, we distinguish different degrees of emissivity. For example, the majority of the spectra would correspond to ``mild'' Be stars, as the emission is only present in H$\alpha$ \cite{GrayCorbally2021}. For the candidates YSG\_28\_55 YSG\_7\_63, YSG\_88\_20, and YSG\_298\_42, the H$\beta$ emission appears strong enough to fill in the absorption line. For two candidates YSG\_103\_-1 and YSG\_110\_-21, H$\beta$ also appears in emission, with the last one also showing emission for the Paschen series.

Because all Be stars show variations on short and long timescales, it is not unsurprising that our variability based method has selected heavily extincted Be stars as possible candidates. Nevertheless, while Be stars appear non-variable at IR wavelengths  \citep{Gehrz1974ApJ}, the NEOWISE data for our candidates shows a frequent correlation between variability in the optical and mid-IR. Although the specific modeling of the IR SEDs for our candidates is outside of the scope for this work, one plausible explanation for our sources would be the presence of extended dusty disks, capable of reprocessing the optical light. One possible origin for such disks would be mass transfer from a nearby companion.

\subsubsection{Herbig Ae/Be stars}

One alternative classification for at least some of our candidates is that of a pre-main-sequence (PMS) A- or B-type star \citep{Herbig1960ApJS}. These young 2$-$20\,\Msun stars appear to be in the HG, as they are still gravitationally contracting towards their final location in the MS. Their usual association with dark star-formation clouds would also imply large values of line-of-sight extinction, which would make them appear redder and hence susceptible to be picked up by our progenitor selection method. Other characteristics of such stars are Balmer emission lines and IR excess. In contrast to Be stars, where the IR is created by free$-$free and free$-$bound emission in a ionized disk, Herbig Ae/Be stars have large and dusty accretion disks. These stars are highly variable, with major variability episodes related to sudden drops in the optical bands caused by increased column density (UX Ori variables), and slow brightenings or fadings, possibly related to variable accretion rate \citep[FU Ori outbursts;][]{HartmannKenyon1996ARAA} or variations in the circumstellar extinction \citep{Waters1998ARAA}. In addition, some variations may be caused by mass accretion, as 70\% of Herbig Ae/Be stars were found to be in binaries \citep{Baines2006MNRAS}.

Nevertheless, visual inspection of the location for our candidates in PanSTARRS \citep{Chambers_2016} and H$\alpha$ IGAPS images \citep{mongui_2020} did not reveal obvious association with dark stellar clouds or nebulosities in the vicinity of the sources, making the PMS explanation less likely. In addition, the accretion disks of young Herbig Ae/Be contain hot circumstellar dust with temperatures of $\sim$1500\,K, which becomes easily detectable at wavelengths longer than $2$\,$\mu$m \citep{Waters1998ARAA}. Provided that for all our candidates the IR excess is only detectable at mid-IR wavelengths, our systems would be more consistent with young Vega-type MS stars that still preserve their circumstellar disks \citep{Walker1988PASP}. However, no considerable variability is expected for these sources, ruling them out as possible explanation for our candidates.

\subsubsection{Binaries with compact companions}

Among our candidates, YSG\_103\_-1 (V520\,Mon) was detected as a soft variable X-ray source, becoming a good candidate for a binary system with a compact companion. Its highly variable light curve in the DASCH archive shows that the source had a rather erratic behaviour from 1900 until 1980, with a likely major episode of increased accretion around 1960. Although at a smaller scale, most of our candidates also show non-periodic variability in historic observations. Provided that the range of the historic variations is consistent with the changes detected from ZTF data, we deem unlikely for our sources to be LRN precursors with an imminent outburst. However, the origin of their variability is of considerable interest.

For binaries in non-circular orbits, enhanced mass transfer at periastron can fuel higher accretion rates, capable of generating the enhanced luminosities observed in our data \citep{Davis_2013}. Moreover, changes in the properties of the accretion disk, such as creation and destruction of hot spots, can further contribute to the source's variability.

\subsubsection{Recent merger products}
One final possibility would be that the observed sources correspond to recent stellar mergers. The previous unstable mass transfer episodes in the binary would explain the existence of the circumstellar dusty disk that we detect for our candidates. After coalescence, the merger product is expected to show a considerable increase in its magnetic activity \citep{SokerTylenda2007MNRAS,Schneider2019Natur}, due to a stellar dynamo effect, generated by the convection region created in the extended envelope of the merger product, created by its fast rotation. Magnetorotational instabilities could generate turbulence in the accretion disk, which can power some of the luminosity episodes seen in our sources. This hypothesis could be tested using X-ray observations, as there is an empirical connection between magnetic flux in the stellar surface and X-ray emission \citep{Pevtsov2003ApJ}. Unfortunately, our cross-match with X-ray catalogues only allowed the identification of the source YSG\_103\_-1 (V520\,Mon).

\section{Conclusion}\label{sec:Conclusion}

In this work, we developed a method to identify Galactic LRN precursors by first statistically selecting 54347 LRN progenitor candidates (YSG and YG) from CMDs constructed from Gaia DR2 and EDR3. Analysis of the progenitor candidates' ZTF time-domain light curves resulted in the identification of 21 LRN precursor candidates, 20 of which were follow-up with optical low-resolution spectra. 

The candidates found in our study are broadly consistent with low-mass 2$-$5\,\Msun emission line stars with spectral types Be and Ae, although larger masses are not discarded, as donor stars in semi-stable mass transfer can decrease their luminosity up to an order of magnitude \citep{Klencki2021AA}. Because of their intrinsic reddening, the sources are located in a sparsely populated area of the HRD, and hence were selected by our LRN progenitor method as potential YG and YSG candidates. The non-periodic variability in our sources is also seen in historic DASCH data, which rules them out as mass transferring binaries 5$-$10\,years before their merger. However, the nature of variability in these sources is intriguing. 

Although high resolution spectroscopy would be required for accurate typing, the candidates share several characteristics. Most of them show emission in the H$\alpha$ and even H$\beta$ line, but their IR excess and variability are inconsistent with the ionized disks in typical Be stars. Their lack of NIR excess and location away from dark clouds or nebulae makes them unlikely to be Herbig Ae/Be stars. Therefore, we hypothesize that they could potentially be either mass-transferring binaries with compact companions surrounded by dusty circumstellar disks or magnetically active stellar merger remnants. X-ray follow-up observations, IR excess modelling, and high resolution spectroscopy would be required to establish a definitive answer to this question.

Future research on Galactic LRN precursors will benefit from the Gaia DR3 \citep{Gaia_DR3}, which will provide us with the most complete Galactic survey of stars yet. The release of astrophysical parameters, such as effective temperatures and extinction values ($\rm{A_g}$) obtained from BP/RP spectra for 470 million objects will allow us to make our progenitor selection from an extinction-corrected HR diagram, minimizing contamination from reddened sources and including more intrinsically redder stars in our sample. Furthermore, H$\alpha$ emission in 235 million stars will be provided along with BP/RP spectra for 219 million objects. This data will allow us to develop complementary selection methods based on the spectral features of the sources, and use the presence of emission lines to prioritize follow-up with higher resolution spectrographs.

\section*{Acknowledgments}
This work is part of the research programme VENI, with project number 016.192.277, which is (partly) financed by the Netherlands Organisation for Scientific Research (NWO). DJ acknowledges support from the Erasmus+ programme of the European Union under grant number 2020-1-CZ01-KA203-078200. PJG  and OM are supported by NRF SARChI grant 111692

The DASCH project at Harvard is grateful for partial support from NSF grants AST-0407380, AST-0909073, and AST-1313370; which should be acknowledged in all papers making use of DASCH data.

We acknowledge the one-time gift of the Cornel and Cynthia K. Sarosdy Fund for DASCH, and thank Grzegorz Pojmanski of the ASAS project for providing some of the source code on which the DASCH web-interface is based.

The ongoing AAVSO Photometric All-Sky Survey (APASS) has improved DASCH photometric calibration and is funded by the Robert Martin Ayers Sciences Fund.

The Liverpool Telescope is operated on the island of La Palma by Liverpool John Moores University in the Spanish Observatorio del Roque de los Muchachos of the Instituto de Astrofisica de Canarias with financial support from the UK Science and Technology Facilities Council.

This work made extensive use of Python, specifically the packages: NumPy \citep{harris_2020}, Astropy \citep{Astropy_2013, Astropy_2018}, Matplotlib \citep{Hunter_2007}, and Astroquery \citep{Ginsburg_2019}.

This work has made use of data from the Asteroid Terrestrial-impact Last Alert System (ATLAS) project. The Asteroid Terrestrial-impact Last Alert System (ATLAS) project is primarily funded to search for near earth asteroids through NASA grants NN12AR55G, 80NSSC18K0284, and 80NSSC18K1575; byproducts of the NEO search include images and catalogs from the survey area. This work was partially funded by Kepler/K2 grant J1944/80NSSC19K0112 and HST GO-15889, and STFC grants ST/T000198/1 and ST/S006109/1. The ATLAS science products have been made possible through the contributions of the University of Hawaii Institute for Astronomy, the Queen’s University Belfast, the Space Telescope Science Institute, the South African Astronomical Observatory, and The Millennium Institute of Astrophysics (MAS), Chile. This work has made use of data obtained with telescopes and instruments supported by the South African National Research Foundation (NRF).

\section*{Data availability}
All data are available upon reasonable request to the corresponding author.

\bibliographystyle{mnras}
\bibliography{ref}

\bsp	
\label{lastpage}
\end{document}